\documentclass[10pt,superscriptaddress,english,showpacs,longbibliography,nofootinbib,floatfix]{revtex4-1}
\pdfoutput=1
\usepackage{amsmath}
\usepackage{hyperref}
\usepackage{xurl}
\usepackage{amssymb}
\usepackage{titletoc}
\usepackage{multirow}
\usepackage{enumitem}
\usepackage{soul}
\usepackage{graphicx}
\usepackage{amssymb}
\usepackage[super]{nth}
\usepackage[normalem]{ulem} 
\usepackage{longtable}
\usepackage{dsfont}

\newcommand{\Pso}{P$_{\textrm{\footnotesize{SO}}}$}
\newcommand{\Ps}{P$_{\textrm{\footnotesize{S}}}$}

\newcommand{\Pnt}{P$_{\textrm{\footnotesize{NT}}}$}

\newcommand{\Psog}{P$_{\textrm{\footnotesize{SOG}}}$}
\newcommand{\Pe}{P$_{\textrm{\footnotesize{E}}}$}

\newcommand{\refspgenpos}{S1}

\newcommand{\refexception}{S1.4}

\newcommand{\refelementary}{S1.5}

\newcommand{\tablenocoplanar}{S2}

\newcommand{\tablecoplanar}{S3}

\newcommand{\spinsplitnocoplanar}{S4}

\newcommand{\spinsplitnocoplanarTR}{S5}

\newcommand{\spinsplitcoplanar}{S6}

\newcommand{\spinsplitcoplanarTR}{S7}

\begin{document}

\title{Spin point group symmetry and classification of non-relativistic spin splitting in non-collinear magnetic structures: Identification of high-order spin splitting types ($\ell$=5,7, and 9)}

\author{Luis Elcoro}
\email{luis.elcoro@ehu.eus}
\affiliation{Department of Physics, Faculty of Science and Technology, UPV/EHU, Bilbao, Spain}

\author{Jesus Etxebarria}
\email{j.etxeba@ehu.eus}
\affiliation{Department of Physics, Faculty of Science and Technology, UPV/EHU, Bilbao, Spain}

\author{J. Manuel Perez-Mato}
\email{jm.perezmato@gmail.com}
\affiliation{Faculty of Science and Technology, UPV/EHU, Bilbao, Spain}

\author{Emre S. Tasci}
\affiliation{Department of Physics Engineering, Hacettepe University, 06800 Ankara, Turkey}
\affiliation{Materials Genome Institute, Shanghai University, 200444 Shanghai, China}

\begin{abstract}
	A comprehensive study of the possible types of non-relativistic spin splitting of electronic bands in coplanar and non-coplanar magnetic structures is presented on the basis of spin-group theory. As a first step, we enumerate and tabulate all possible non-equivalent spin point groups (SpPGs) which can be expressed as a direct product of a nontrivial part and a spin-only group limited to be the intrinsic (trivial) one, or augmented by the time-reversal operation. This tabulation, which includes the listing of symmetry operations for each group, is now available as an online database \mbox{(\emph{SPGENPOS})} in the Bilbao Crystallographic Server. A total of 1249 nonequivalent SpPGs are listed in \mbox{\emph{SPGENPOS}}. This extends previous enumerations, in which the possible presence of time reversal in the magnetic point group was not taken into account, thus overlooking the full SpPG symmetry associated with the numerous magnetic structures which have a magnetic space group of type IV. For each of the listed coplanar and non-coplanar SpPGs, the spin-splitting that is symmetry allowed is analyzed in detail using the program \mbox{\emph{STENSOR}} also in the Bilbao Crystallographic Server. Except for the SpPGs that include the operation $\overline{1}^{\prime}$, i.e., the combined operation of time reversal and space inversion, all other coplanar and non-coplanar SpPGs allow spin splitting at some order in a power expansion of the electron wave vector components. We find that, depending on the SpPG, spin-splitting terms can appear with the lowest-order monomials ranging from $\ell$ =0 to 9, with the exception of $\ell$=8. This contrasts with the collinear case, where the lowest order is not higher than $\ell$=6, and where the presence of time-reversal symmetry forbids any spin splitting. For the newly identified spin textures with powers $\ell$=5, 7, and 9, which are possible in some noncentrosymmetric SpPGs, the functional form of the spin splitting in terms of the components of the crystal momentum is given. One example of a real material, LaMnAu$_5$, showing $\ell$=5 spin splitting is identified. \end{abstract}

\maketitle

\section{Introduction}
	\label{sec:introduction}
	The development of magnetic structures whose electronic bands exhibit non-relativistic spin splitting is a major objective in the field of magnetic materials science, owing to its potential relevance for possible spintronics applications \cite{Hayami2019,Yuan2020,Mazin2021}. A major breakthrough toward this target has been the introduction of the concept of altermagnetism \cite{Smejkal2022a,Smejkal2022b,Mazin2022,Krempasky2024,Bhowal2024}, differentiating a class of compensated collinear magnetic states that have magnetic point groups (MPGs) lacking both the operation of time-reversal (TR) symmetry $1^{\prime}$ and TR combined with space inversion $\overline{1}^{\prime}$. These systems exhibit a remarkable even-parity spin splitting in the absence of spin-orbit coupling (SOC), which enables the occurrence of the anomalous Hall effect and promotes phenomena of potential interest for device applications, such as spin transport driven by applied electric fields or temperature gradients \cite{Gonzalez2021,Cui2023,Han2024}. Within altermagnets, one can further distinguish materials having a spin splitting of so-called $d$, $g$, and $i$-character, indicating a spin splitting near the $\Gamma$ point of the Brillouin zone (${\bf k}$=0) that, at lowest order, can be expressed as a homogeneous polynomial of degree $\ell$ in the components of ${\bf k}$, with $\ell$=2,4,6, respectively. The case $\ell$=0 corresponds to ferromagnetic (FM) materials, which exhibit spin splitting at the center of the Brillouin zone even in the absence of SOC.

Recently, the concept of altermagnetism has been extended to non-collinear noncentrosymmetric magnetic states (Yuan et al. 2021, Hellenes et al. 2024), which allow for non-relativistic spin splittings with $\ell$=1 (type $p$) or $\ell$=3 (type $f$). These magnets are characterized by a spin splitting that changes sign upon inversion of the crystal momentum ${\bf k}$. \citet{Hellenes2024}, and \citet{Priessnitz2026}, pointed out that $p$-wave magnets preserve (TR) as a point-group symmetry operation and, in fact, the search for materials of this type has been formulated under this symmetry constraint. On the other hand, although some works on compensated magnets with non-relativistic spin splittings briefly mention the possibility of systems where the allowed lowest-order spin splitting has $h$-character ($\ell$=5) \cite{Yu2025,Priessnitz2026}, other studies have argued \cite{Ezawa2025} that such states are impossible. Interestingly, by combining low-energy modelling with density functional theory (DFT), \citet{Dsouza2026} have recently proposed that superconductor FeSe should exhibit $h$-wave spin splitting, based on a theoretical crystal structure deduced from DFT calculations. To the best of our knowledge, however, the highest odd-parity value of $\ell$ reported so far in an experimental structure is $\ell$=3. 
	
In this work we present a comprehensive analysis of the possible types of SOC-free spin splitting in non-collinear magnetic structures from the perspective of spin-group symmetry \cite{Chen2024,Jiang2024,Xiao2024}. The application of spin-point-group symmetry allows to resolve in any tensor property which part of the tensor is not necessarily SOC assisted, provided that the spin arrangement does not suffer some lowering of its spin-group symmetry due to SOC \cite{etxebarria2025}. Consequently, we have first established and tabulated all possible  non-equivalent spin point groups (SpPGs) which can be expressed as a direct product of a nontrivial part and a spin-only group limited to be the intrinsic (trivial) one, or augmented by the TR operation. The antitranslations, which combine TR with a translation, are very common, as they are present in the symmetry of all structures with magnetic space groups (MSGs) of type IV, and therefore, their MPG and their SpPG necessarily include TR.  We have then used the program \emph{STENSOR} (\href{https:/cryst.ehu.eus/cryst/stensor.html}{https://cryst.ehu.es/cryst/stensor.html}) \cite{Elcoro2026} to obtain, for all the tabulated coplanar and non-coplanar SpPGs, their symmetry constraints on the possible band spin-splitting around the $\Gamma$ point of the Brillouin zone.

The structure of the paper is as follows. In Section \ref{sec:enumeration} we present and explain the enumeration and tabulation of the SpPGs, including the notation and conventions that will be used. The database \emph{SPGENPOS} (\href{https://www.cryst.ehu.eus/cryst/spgenpos.html}{https://cryst.ehu.es/cryst/spgenpos.html}), available in the Bilbao Crystallographic Server, where this tabulation is hosted, is also described. Section \ref{sec:method} is devoted to outlining the method used to identify the different types of spin splittings, which are then discussed in Section \ref{sec:classification} for the 1127 tabulated non-collinear SpPGs. Finally, in Section \ref{sec:conclusions} we draw our conclusions.
 \section{Enumeration and tabulation of spin point groups. The database \textit{SPGENPOS}}
\label{sec:enumeration}
	The operations $\{U||R\}$ of a SpPG combine a proper or improper spin rotation, represented by the matrix $U$, and a proper or improper spatial rotation, represented by the matrix $R$. The operations $U$ act on the spin components $(u,v,w)$ defined with respect to a spin basis that in general is fully independent of the basis used for the operations $R$ acting on the spatial coordinates $(x,y,z)$. The SpPG associated with a magnetic structure is formed by all pairs $\{U||R\}$ that are present in the operations $\{U||R|{\bf t}\}$ defining the spin space group of the structure, where ${\bf t}$ is a spatial translation.

An SpPG may have spin-only operations of the type $\{U||1\}$, which form the so-called spin-only subgroup denoted by \Pso. The spin space group of most magnetic structures is such that the SpPG, \Ps, can be expressed as the direct product of two subgroups:
\begin{equation}
	\label{equationationdirectproduct}
	\textrm{P}_{\textrm{\footnotesize{S}}}=\textrm{P}_{\textrm{\footnotesize{NT}}}\times\textrm{P}_{\textrm{\footnotesize{SO}}}
\end{equation}
where \Pnt\, is a group enumerated by Litvin in 1977, and termed as “nontrivial” spin point groups, with the idea that the spin-only groups \Pso\, could only be the “trivial” spin-only groups, which are intrinsic to the collinear or the coplanar spin arrangements \cite{Litvin1977}.

However, magnetic structures may have more complex \Pso\, groups if their spin space groups include a spin-translation subgroup \cite{Chen2024}, formed by operations of the type $\{U||1|{\bf t}\}$, with ${\bf t}$ a non-zero translation and $U$ different from the identity. In such cases, additional point group operations $\{U||1\}$ derived from these space group operations must be included in its SpPG. Antitranslations associated with any magnetic structure having an MSG of type IV are operations of this type, and imply the presence of the TR operation $1^{\prime}=\{-1||1\}$ in \Pso. More generally, spin space groups may contain spin-translation operations with spin operations $U\neq1$ giving place to enlarged spin-only subgroups \Pso. In such cases, the validity of equation (\ref{equationationdirectproduct}) can even fail. This corrects the general basis on which the enumeration in \citet{Litvin1977} was done, and our assumption about the general validity of equation (\ref{equationationdirectproduct}) in previous publications \cite{etxebarria2025,Elcoro2026}. A simple example of an SpPG for which equation (\ref{equationationdirectproduct}) does not hold is given in Section \refexception\, of the Supporting Information. In this respect see also Appendix B in \citet{Shinohara2024} and Section \refelementary\, in the Supporting Information. The potential impossibility of a direct product decomposition given by equation (\ref{equationationdirectproduct}) for point groups contrasts with the case of space groups, where the only spin-only subgroups are the intrinsic ones —associated with either the collinearity or the coplanarity—, and where an equation analogous to equation (\ref{equationationdirectproduct}) is always fulfilled.

Thus, when dealing with spin point-group symmetry, there are cases in which SpPGs cannot be generated by applying equation (\ref{equationationdirectproduct}) to one of the 598 so-called “non trivial” SpPGs and the corresponding \Pso. In practice, however, in most of the observed magnetic structures the lattice periodicity either remains the same as that of the paramagnetic phase or is reduced by a factor two, which favors the validity of equation (\ref{equationationdirectproduct}). Therefore the 598 groups enumerated by Litvin can be taken as the set of all possible non-equivalent point groups which lack spin-only operations except the identity. The SpPGs of most of the observed magnetic structures can then be derived from them by taking the direct product with the relevant spin-only subgroup \Pso.

In summary, although equation (\ref{equationationdirectproduct}) is not valid for all mathematically possible SpPGs, it is fulfilled by those of a majority of the observed magnetic structures, and its systematic application generates most of the SpPGs which are relevant in practice. We have therefore enumerated and listed all non-equivalent SpPGs generated from the 598 \Pnt\, groups by applying equation (\ref{equationationdirectproduct}) with the intrinsic \Pso\, subgroup and with this group augmented with TR. If \Pso\, is different from the identity, different \Pnt\, groups multiplied by the same \Pso\, can lead to equivalent SpPGs, and this issue has been taken into account in the enumeration. The resulting SpPGs can be then divided into the following subsets, according to the spin-only group \Pso\, considered (see Section \refspgenpos\, of the Supporting Information for notation):

\begin{enumerate}
	\item Intrinsic collinear \Pso: $^{\infty_w m}1$  (all rotations around the spin direction, which is taken along $w$ in the spin basis, and all mirror planes containing this direction).
	
	This results in 90 collinear SpPGs, isomorphic with the 90 non-gray MPGs \cite{Liu2022}. These SpPGs span all possible spin-point-group symmetries of collinear magnetic structures with MSG of type I or III.
	
	\item Intrinsic collinear \Pso\, with TR: $^{\infty_w/mm}1=\, ^{\infty_w m}1 \times \{\{1||1\},\{-1||1\}\}$.
	
	This results in 32 additional collinear SpPGs (isomorphic with the 32 gray MPGs). These SpPGs span all possible spin-point-group symmetries of collinear magnetic structures with an MSG of type IV.
	
	\item Intrinsic coplanar \Pso: $^{m_w}1$ ($\{\{1||1\}, \{m_w || 1\}\}$).
	
	This results in 253 coplanar SpPGs \cite{Liu2022}. These groups span all possible spin-point-group symmetries of coplanar magnetic structures with MSG of type I or III, provided that its spin space group does not include a spin-translation subgroup. This latter condition is fulfilled by a majority of magnetic structures of type I and III, where the magnetic order maintains the lattice of the paramagnetic phase.
	
	\item Intrinsic coplanar \Pso\, with TR: $^{2_w/m_w}1=\, ^{m_w}1 \times \{\{1||1\},\{-1||1\}\}$.
	
	This results in 114 additional coplanar SpPGs (see Table \tablecoplanar\, in the Supporting Information for a list of equivalences when $1^{\prime}$ is added to the intrinsic coplanar \Pso). These groups span most of the spin-point-group symmetries of coplanar magnetic structures with a MSG of type IV, where the magnetic lattice differs from the paramagnetic one by only a doubling of the primitive unit cell.
	
	\item Intrinsic non-coplanar \Pso: it includes only the identity $\{1||1\}$.
	
	This results in 598 non-coplanar SpPGs coinciding with the 598 \Pnt\, groups. These groups describe all non-coplanar magnetic structures with MSGs of type I or III, provided that their spin space groups do not include some spin-translation subgroups. This latter condition is fulfilled by a majority of the magnetic structures of type I or III, where the magnetic order maintains the lattice of the paramagnetic phase, forbidding any spin-translation group.
	
	\item Intrinsic non-coplanar \Pso\, with TR: $\{\{1||1\},\{-1|1\}\}$.
	
	This results in 162 non-coplanar SpPGs. (see Table \tablenocoplanar\, in the Supporting Information for a list of equivalences when the operation $1^{\prime}$ is added to the nontrivial SpPGs). These groups span most of the possible spin-point-group symmetries of non-coplanar magnetic structures with MSG of type IV, where the magnetic ordering decreases the lattice periodicity with respect to the paramagnetic phase by only a factor 2.
\end{enumerate}

Table \ref{tab:numbergroups} summarizes the classification of the resulting 1249 distinct SpPGs. Detailed tables for each of these groups, including a full list of symmetry operations, are available via the \mbox{\emph{SPGENPOS}} program (\href{https:/cryst.ehu.eus/cryst/spgenpos.html}{https://cryst.ehu.es/cryst/spgenpos.html}) of the Bilbao Crystallographic Server. The SpPGs are identified using both an extension of the Hermann-Mauguin symbols, similar to that used by \citet{Litvin1977}, and also a numerical labeling scheme. A full detailed explanation of the contents of \mbox{\emph{SPGENPOS}} can be found in the Supporting Information. This tabulation of SpPGs available in \mbox{\emph{SPGENPOS}} complements previous enumerations \cite{Liu2022,Schiff2025}, where only SpPGs without TR were considered, so that the spin-point-group symmetries of all magnetic structures with antitranslations in their magnetic symmetry were ignored. As about 40\% of the known commensurate magnetic structures are of this type \cite{Gallego2016}, this left a considerable number of magnetic materials out of the spin-point-group framework. 

\begin{table}[h]
	\caption{\label{tab:numbergroups} Number of SpPGs in \mbox{\emph{SPGENPOS}} classified according to the six types of spin-only subgroups considered in this work.}
	\begin{tabular}{|c|c|c|c|c|c|c|}
		\hline
		collinear&collinear&coplanar&coplanar&non-coplanar&non-coplanar&Total\\
		without TR&with TR&without TR&with TR&without TR&with TR&\\
		\hline
		90&32&253&114&598&162&1249\\
		\hline
	\end{tabular}
\end{table}

Note that the SpPGs tabulated in \mbox{\emph{SPGENPOS}}, as stressed above, do not exhaust all possible SpPGs, although they cover by far the most common cases. To these, one must add the SpPGs that cannot be expressed in the form of equation (\ref{equationationdirectproduct}) as well as those that, despite satisfying equation (\ref{equationationdirectproduct}), have a \Pso\, different from those considered in this tabulation. Although relatively uncommon, there are known examples of real materials where this happens (see sections \refexception\, and \refelementary\, in the Supporting Information).  \section{Method for the spin-splitting analysis}
\label{sec:method}
	The spin splitting of the energy bands can be considered to arise from an effective Zeeman field ${\bf B}({\bf k})$, which depends on the wave vector ${\bf k}$ and the band index. For a particular band, the direction of the spin polarization and the magnitude of the non-relativistic spin splitting at a given ${\bf k}$ are determined by the direction and magnitude of ${\bf B}({\bf k})$ at that point \cite{Radaelli2024}. The general features of the ${\bf B}({\bf k})$ field can be determined using symmetry arguments, by imposing that, in the absence of SOC, it is constrained to remain invariant under the SpPG operations $\{U||R\}$. ${\bf B}$ transforms as a spin magnetization, i.e., $\{U||R\}{\bf B}=U {\bf B}$, whereas the wave vector becomes $\{U||R\}{\bf k}=\det(U) R {\bf k}$, where $\det(U)$ denotes the determinant of the matrix $U$ \cite{etxebarria2025,Elcoro2026}. Therefore, the invariance condition of ${\bf B}$ under the SpPG is written as
\begin{equation}
	\label{eq:transfB}
	{\bf B}({\bf k})=U{\bf B}(\det(U)R{\bf k})
\end{equation}
Using equation (\ref{eq:transfB}), we can derive in a straightforward manner some properties for the spin splittings, which are broadly applicable, as they originate from operations such as $1^{\prime}$, $\overline{1}$, and/or $\overline{1}^{\prime}$ , which are common operations of both the SpPGs and corresponding MPGs, or from spin-only operations that are present in all collinear or coplanar SpPGs. We briefly outline some of these simple general points —some of them rather well known— that follow directly from equation (\ref{eq:transfB}), taking into account that the MPG of a magnetic structure is always a subgroup of its SpPG, and, therefore, the MPG operations are also SpPG operations:

\renewcommand{\theenumi}{\roman{enumi}}\begin{enumerate}
	\item\label{enu:inversion} If the space inversion operation $\{1||\overline{1}\}$ is present in the MPG, i.e., the group is centrosymmetric, then ${\bf B}({\bf k})={\bf B}(-{\bf k})$, and the spin splitting, relativistic or not, can only be of even parity.
	
	\item If the TR operation $1^{\prime}= \{-1||1\}$ is present in the MPG, then ${\bf B}({\bf k})=-{\bf B}(-{\bf k})$, and the spin splitting, relativistic or not, can only be of odd parity. Considering point (\ref{enu:inversion}) above, this means that centrosymmetric groups with TR imply no spin-splitting in any case (see point (\ref{enu:inversionTR}) below).
	
	\item\label{enu:inversionTR}  The presence of the operation $\overline{1}^{\prime}=\{-1||\overline{1}\}$ in the MPG (often denoted as PT symmetry), forbids any spin splitting in any case since equation (\ref{eq:transfB}) implies ${\bf B}({\bf k})=-{\bf B}({\bf k})$, forcing ${\bf B}=0$. The same operation also prevents any macroscopic magnetization \cite{etxebarria2025}. This is the case of the so-called non-spin-split pure antiferromagnets \cite{Liu2026}.
	
	\item\label{enu:TR} Nonmagnetic structures are symmetric for TR. Consequently, spin splitting is only permitted in noncentrosymmetric structures, since inversion symmetry would imply the presence of the operation $\overline{1}^{\prime}$, which leads to ${\bf B}=0$, as indicated in point (\ref{enu:inversionTR}). Moreover, in these noncentrosymmetric nonmagnetic structures, the spin splitting can arise exclusively due to SOC, as the SpPGs of such structures also contain operations of the type $\{U||1\}$ for arbitrary rotations $U$, which enforce ${\bf B}=0$ in the absence of SOC.
	
	\item For collinear structures the spin-only group, $^{\infty_{\bf n}m}1$, already ensures that ${\bf B}({\bf k})$ is parallel to the spin direction. Choosing the spin direction along the third axis in the spin plane ($w$-axis), then $\{\infty_w||1\}$ in equation (\ref{eq:transfB}) directly yields $B_u=B_v=0$, where $u$, $v$, and $w$ form a mutually orthogonal set of axes in the spin space. The constraint of the spin-only operation $\{m||1\}$, with $m$ parallel to the $w$-axis, also forces $B_w({\bf k})=B_w(-{\bf k})$. This means that collinear structures can only have spin splittings of even parity in ${\bf k}$. If, in addition, the point group of the collinear structure includes TR, then from point (\ref{enu:TR}) we have that $B_w({\bf k})=-B_w(-{\bf k})$, which leads to ${\bf B}=0$. Therefore, collinear structures whose point group includes the time reversal operation do not allow non-relativistic spin splitting and thus any experimentally observed splitting must be SOC assisted.
	
	\item For coplanar structures the components of ${\bf B}$ within the spin plane are even in ${\bf k}$, whereas the component perpendicular to the plane is odd in ${\bf k}$. Indeed, choosing the spin plane perpendicular to the $w$-axis in the spin space, the coplanar spin-only operation $\{m_w||1\}$ imposes ${\bf B}({\bf k})=m_w {\bf B}(-{\bf k})$. As a result, $B_u({\bf k})$ and $B_v({\bf k})$ must be even functions of ${\bf k}$, while $B_w({\bf k})=- B_w(-{\bf k})$. If TR symmetry is also present, then $B_u=B_v=0$, the electronic spin polarization is perpendicular to the spin plane, and the spin splitting is necessarily of odd-parity.
\end{enumerate}

Having recapitulated these general properties for the effective spin splitting field ${\bf B}$, we now turn to the detailed analysis of the possible types of spin splitting. These types can be characterized by the minimum degree $\ell$ that is symmetry allowed in a power expansion of ${\bf B}$ in terms of the components of the electron wave vector. Thus, we write ${\bf B}$ as a Taylor series around ${\bf k}=0$,

\begin{equation}
	\label{eq:Bfield}
	B_i=\sum_{n=0}^{\infty}T^{(n)}_{i,\alpha_1,\alpha_2,\ldots,\alpha_n}k_{\alpha_1}k_{\alpha_2}\ldots k_{\alpha_n}
\end{equation}

where $k_{\alpha}$ are the components of the wave vector ${\bf k}$, and $T^{(n)}$ is a tensor of rank $n+1$ symmetric in indices $\alpha_i$. The transformation laws of these tensors $T^{(n)}$ under a symmetry operation $\{U||R\}$ are deduced from those of ${\bf B}$ and ${\bf k}$. The components of the transformed tensors $T^{'(n)} =\{U||R\} T^{(n)}$ are given by

\begin{equation}
	\label{eq:Ttensoreven}
	T^{\prime(n)}_{j,\beta_1,\beta_2,\ldots,\beta_n}=U_{ji}R_{\beta_1,\alpha_1}R_{\beta_2,\alpha_2}\ldots R_{\beta_n,\alpha_n}T^{(n)}_{i,\alpha_1,\alpha_2,\ldots,\alpha_n}\hspace{1cm}(n\textrm{ even})
\end{equation}

\begin{equation}
	\label{eq:Ttensorodd}
	T^{\prime(n)}_{j,\beta_1,\beta_2,\ldots,\beta_n}=\det(U)U_{ji}R_{\beta_1,\alpha_1}R_{\beta_2,\alpha_2}\ldots R_{\beta_n,\alpha_n}T^{(n)}_{i,\alpha_1,\alpha_2,\ldots,\alpha_n}\hspace{1cm}(n\textrm{ odd})
\end{equation}

The transformation laws (\ref{eq:Ttensoreven}) and (\ref{eq:Ttensorodd}) can be compactly expressed using the generalized Jahn symbol for the transformation properties of a tensor \cite{etxebarria2025}. This notation combines the letters M and V with the modifiers $e$ and $a$. Under an operation $\{U||R\}$, the tensor transforms according to the labels in its Jahn symbol: matrix $R$ is used for V, while matrix $U$ is used for M. The letters $e$ and $a$ specify whether a sign change occurs when $R$ or $U$, respectively, represent improper operations. Thus, $e$ indicates that the tensor is even under spatial inversion, and $a$ indicates that it is odd under TR. Square or curly brackets are added sometimes to denote symmetry or anti-symmetry between pairs of indices (if $n\geq2$). Accordingly, ${\bf B}$ is of type M, ${\bf k}$ of type $a$V, whereas $T^{(n)}$ is of type M[V$^n$] when $n$ is even, and of type $a$M[V$^n$] when $n$ is odd.

Equations (\ref{eq:Ttensoreven}) and (\ref{eq:Ttensorodd}) allow for the systematic identification of the type of non-relativistic spin splitting permitted by each SpPG. The spin-splitting type is simply determined by the smallest index $\ell$ for which the tensor $T^{(\ell)}$ is nonzero. If $\ell=0$, the system is FM ($s$-wave); $\ell=1$ corresponds to a $p$-wave magnet; $\ell=2$ corresponds to a $d$-wave magnet, and so on. Note that although the “wave type” has been here defined by the lowest value of $\ell$ —which usually gives the main contribution— this does not preclude the existence of higher-order non-vanishing tensors $T^{(n)}$ ($n>l$). It is also possible that all $T^{(n)}$ vanish, i.e., that the spin splitting is zero. As mentioned above this arises in SpPGs that include the $\overline{1}^{\prime}$ operation and corresponds to the non-spin-split pure antiferromagnets \cite{Liu2026}. In these systems neither macroscopic magnetization nor spin-splitting is allowed, even in the presence of SOC, since the $\overline{1}^{\prime}$ operation also belongs to the corresponding MPG of the structure.  \section{Classification of non-relativistic spin splittings}
\label{sec:classification}
	\renewcommand{\theenumi}{\alph{enumi}}\begin{enumerate}
	\item Coplanar SpPGs.
	
	For each coplanar SpPG, the forms of the tensors $T^{(\ell)}$ allowed by the corresponding symmetry were obtained using the program \emph{STENSOR} (\href{https:/cryst.ehu.eus/cryst/stensor.html}{https://cryst.ehu.es/cryst/stensor.html}) \cite{Elcoro2026}, a computational tool integrated into the Bilbao Crystallographic Server and designed for the automatic calculation of symmetry-adapted tensors under SpPG symmetry. The program is directly accessible from \emph{SPGENPOS}. A summary of the results is presented in Table \ref{tab:coplanar}, which lists the number of SpPGs for each spin splitting type. For comparison, the well-known results for collinear SpPGs are also shown in Table \ref{tab:collinear}. As can be seen, among the coplanar SpPGs analyzed, there exist examples that allow non-relativistic spin splitting with waves ranging from $\ell=0$ to $\ell=9$, with the exception of $\ell=8$. This constitutes the most noticeable difference with respect to the collinear case, where $\ell=0,2,4$, and 6 span all possible minimal orders in the spin splitting (Table \ref{tab:collinear}). In particular, the $h$-wave ($\ell=5$) occurs for 5 coplanar SpPGs, all of them tetragonal with TR and necessarily noncentrosymmetric. This explicitly shows that the argument linking $h$-waves to 5-fold rotational symmetry is incorrect in \citet{Ezawa2025}. Likewise, $j$-wave splitting ($\ell=7$) appears for 5 hexagonal (noncentrosymmetric) SpPGs with TR, and $\ell$-wave splitting ($\ell=9$) for 4 cubic (noncentrosymmetric) SpPGs with TR. This also contrast with the collinear case, where spin splitting is forbidden if the point group includes TR.
	
	\begin{table}[h]
		\caption{\label{tab:coplanar} Number of coplanar SpPGs for each type of non-relativistic spin splitting waves.}
		\begin{tabular}{|c|c|c|c|c|c|c|c|c|c|c|c|c|}
			\hline
			Type&No splt.&$s$ (FM)&$p$&$d$&$f$&$g$&$h$&$i$&$j$&$k$&$\ell$&Total\\
			 of splt.&(with $\overline{1}^{\prime}$)&$\ell=0$&$\ell=1$&$\ell=2$&$\ell=3$&$\ell=4$&$\ell=5$&$\ell=6$& $\ell=7$&$\ell=8$&$\ell=9$&\\
			\hline
			without TR&46&90&36&46&17&12&0&6&0&0&0&253\\
			\hline
			with TR&29&0&37&0&34&0&5&0&5&0&4&114\\
			\hline
		\end{tabular}
	\end{table}

\begin{table}[h]
	\caption{\label{tab:collinear} Number of collinear SpPGs for each type of non-relativistic spin splitting waves. }
		\begin{tabular}{|c|c|c|c|c|c|c|}
	\hline
	Type&No splt.&$s$ (FM)&$d$&$g$&$i$&Total\\
	of splt.&&$\ell=0$&$\ell=2$&$\ell=4$&$\ell=6$&\\
	\hline
	without TR&21 (with $\overline{1}^{\prime}$)&32&15&15&7&90\\
	\hline
	with TR&32&0&0&0&0&32\\
	\hline
\end{tabular}
\end{table}

Table \ref{tab:coplanarfunctions} provides a more detailed account of the SpPGs that permit these “exotic” spin splitting types. The expressions for the splittings with these symmetries are also indicated (except for a constant) in the same Table. Figure \ref{fig:waves} shows schematically the splitting for $h$, $j$ and $l$ waves close to ${\bf k}=0$. The shape of the curves is complicated as corresponds to the high $\ell$ values. The complete results for the 367 coplanar SpPGs are given in the Supporting Information (Tables \spinsplitcoplanar\, and \spinsplitcoplanarTR). For even-parity modes the electronic polarization lies always on the spin plane ($B_w=0$) while for odd-parity splittings it is perpendicular ($B_u=B_v=0$). This contrasts with the collinear case, where the polarization is always along the spin direction.

	\begin{table}[h]
	\caption{\label{tab:coplanarfunctions} Coplanar SpPGs showing spin splittings of $h$, $j$, and $\ell$-wave type. The suffix $P1^{\prime}$ in the group labels indicates that \Pso\, is coplanar and includes TR symmetry. In all cases the spin plane is taken perpendicular to the $w$-axis in the spin basis. The spin splittings for the $h$, $j$ and $\ell$-waves are given up to a constant. For $\ell=7$, the reference frame used to express the components of the wave vector is an orthonormal setting with axes parallel to ${\bf a}$, ${\bf a}+2{\bf b}$, and ${\bf c}$. For $\ell=5$ and 9, an orthonormal basis with axes parallel to ${\bf a}$,${\bf b}$, and ${\bf c}$ is used.}
	\begin{tabular}{|c|c|c|c|c|c|c|}
		\hline
		Type of splt.&SpPGs&Spin splitting\\
		\hline
		\multirow{5}{*}{$h$ ($\ell=5$)}&$^{1}4^{1}2^{1}2.P1^{\prime}$ (12.130.65.P1')&\multirow{5}{*}{$B_w=k_1k_2k_3\left(k_1^2-k_2^2\right)$}\\
		&$^{1}4^{2_{100}}m^{2_{100}}m.P1^{\prime}$ (13.147.76.P1')&\\
		&$^{4_{001}}4^{2_{100}}m^{2_{110}}m.P1^{\prime}$ (13.159.80.P1')&\\
		&$^{2_{100}}\overline{4}^{1}2^{2_{100}}m.P1^{\prime}$ (14.169.87.P1')&\\
		&$^{1}4/^{2_{100}}m^{2_{100}}m^{2_{100}}m.P1^{\prime}$ (15.199.102.P1')&\\
		\hline
		\multirow{5}{*}{$j$ ($\ell=7$)}&$^{1}6^{1}2^{1}2.P1^{\prime}$ (24.338.155.P1')&\multirow{5}{*}{$B_w=k_1k_2k_3\left(3k_1^2-k_2^2\right)\left(k_1^2-3k_2^2\right)$}\\
		&$^{1}6^{2_{100}}m^{2_{100}}m.P1^{\prime}$ (25.394.181.P1')&\\
		&$^{3_{001}}6^{2_{100}}m^{2_{110}}m.P1^{\prime}$ (25.406.185.P1')&\\
		&$^{2_{100}}\overline{6}^{2_{100}}m^{1}2.P1^{\prime}$ (26.419.193.P1')&\\
		&$^{1}6/^{2_{100}}m^{2_{100}}m^{2_{100}}m.P1^{\prime}$ (27.453.209.P1')&\\
		\hline
		\multirow{4}{*}{$\ell$ ($\ell=9$)}&$^{2_{100}}\overline{4}^{1}3^{2_{100}}m.P1^{\prime}$ (31.552.236.P1')&\multirow{4}{*}{$B_w=k_1k_2k_3\left(k_1^2-k_2^2\right)\left(k_2^2-k_3^2\right)\left(k_3^2-k_1^2\right)$}\\
		&$^{2_{100}}\overline{4}^{3_{001}}3^{2_{100}}m.P1^{\prime}$ (31.555.237.P1')&\\
		&$^{1}4^{1}3^{1}2.P1^{\prime}$ (30.559.238)&\\
		&$^{2_{100}}m^{2_{100}}\overline{3}^{2_{100}}m.P1^{\prime}$ (32.574.248.P1')&\\
		\hline
	\end{tabular}
\end{table}

\begin{figure}[ht]
	\centering
	\includegraphics[width=0.8\textwidth]{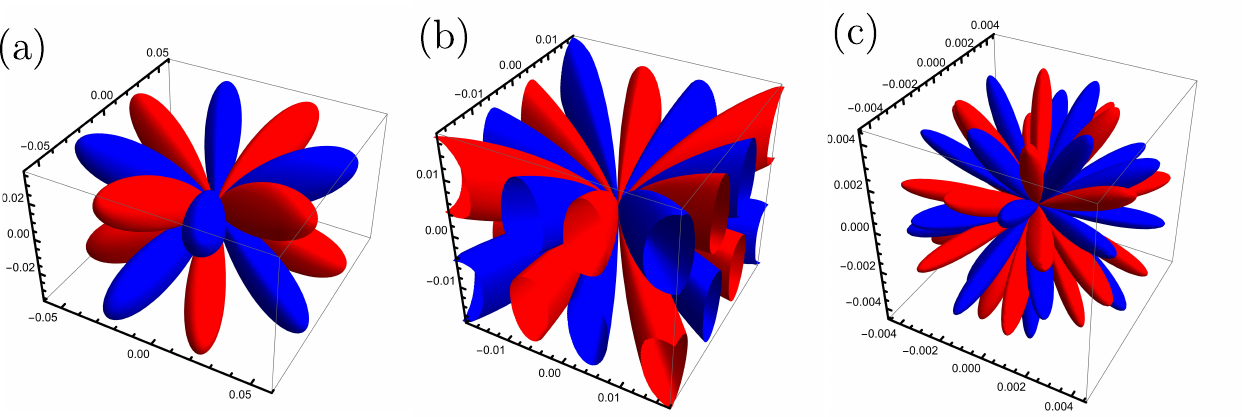}
	\caption{\label{fig:waves}Schematic representation of the SOC-free spin splittings near the center of the Brillouin zone for the (a) $h$-wave , (b) $j$-wave and (c) $\ell$-wave in coplanar magnets. Red (blue) curves indicate positive (negative) splittings. The electronic polarization is perpendicular to the spin plane. The components of the wave vector are in arbitrary units.}
\end{figure}

It is also noteworthy that the lowest order splitting can be odd with low values of $\ell$ (1 and 3) even in the absence of $1^{\prime}$. In fact, as indicated in Table \ref{tab:coplanar}, there is a considerable number of examples where this occurs (36 cases for the $p$-wave and 17 for the $f$-wave). This implies that the TR operation $1^{\prime}$ is not required to suppress the lower-order even tensors ($T^{(0)}$ for the $p$-wave, $T^{(0)}$ and $T^{(2)}$ for the $f$-wave), as these tensors can be constrained to be null by other symmetry operations. Evidently, the inclusion of TR symmetry significantly increases the number of cases with odd-parity splittings, as it automatically suppresses all $T^{(n)}$ with even $n$. In particular, high $\ell$ waves ($h$,$j$,$\ell$) only happen if TR is present.

With the exception of the group $^{4_{001}}4^{2_{100}}m^{2_{110}}m.P1^{\prime}$ (13.159.80.P1') the SpPGs listed in Table \ref{tab:coplanarfunctions} share the general feature that the symmetry of spin space is much lower than the lattice symmetry. The case of $^{4_{001}}4^{2_{100}}m^{2_{110}}m.P1^{\prime}$ is completely different since, except for the coplanar intrinsic spin-only operation $\{m_{001}||1\}$, it corresponds exactly to an MPG, namely $4mm.1^{\prime}$ (N. 13.2.45). Motivated by this coincidence, we carried out a search in MAGNDATA, where the MPGs of all listed compounds are available, with the aim of identifying a real material showing $h$-wave spin splitting. There is only one entry in the entire database possessing this MPG and also being coplanar: LaMnAu$_5$ (entry 1.839), which indeed has $\ell=5$. The structure of this magnetic phase, as determined by powder neutron diffraction \cite{Siebeneichler2024}, is shown in Figure \ref{fig:LaMnAu5}. The SpPG of this structure is therefore either $^{4_{001}}4^{2_{100}}m^{2_{110}}m.P1^{\prime}$ or higher, and one can then expect, according to table \ref{tab:coplanarfunctions}, that its non-relativistic spin-splitting type must be either $\ell=5$ or higher. It can be checked, however, that the SpPG symmetry is indeed higher than simply $4mm.1^{\prime}\times\,^{m_{001}}1$. Introducing the structural data into \emph{FindSpinGroup} \cite{Yu2026}, the spin space group results to be 129.23.2.1.P in the notation of \citet{Chen2024}, or in the generalized Hermann-Mauguin notation:

\begin{figure}[ht]
	\centering
	\includegraphics[width=0.8\textwidth]{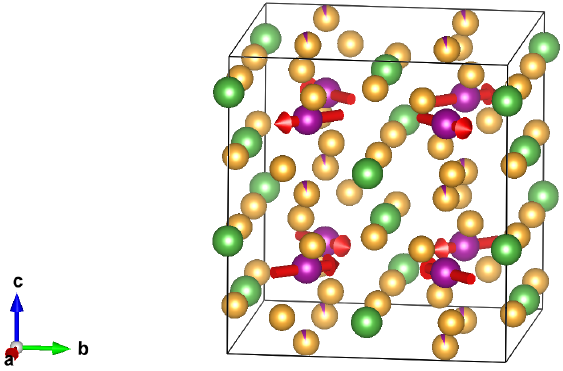}
	\caption{\label{fig:LaMnAu5}Magnetic structure of LaMnAu$_5$. Yellow, green and purple spheres represent Au, La and Mn atoms respectively. Magnetic moments are indicated by arrows.}
\end{figure}

\begin{equation}
	P\,^{4_{001}}4/\,^{2_{110}}n\,^{2_{100}}m\,^{2_{110}}m|(2_{001},2_{001},2_{001})\,^{m_{001}}1
\end{equation}

The SpPG corresponding to this spin space group is generated by adding the operation $\{2_{110}||m_{001}\}$ to the MPG mentioned above. This SpPG does not appear in Table \ref{tab:coplanarfunctions}, because this group cannot be expressed as the direct product of one of the 598 nontrivial point groups and a spin-only group (see Section \refexception\, in the supporting information). This higher symmetry does not modify, however, the final results for the spin splitting type. LaMnAu$_5$ therefore constitutes a real example of an experimental structure that has a SpPG symmetry with an $h$-wave non-relativistic spin splitting.

\item Non-coplanar SpPGs.

Table \ref{tab:non-coplanar} shows the number of non-coplanar SpPGs for each spin splitting type. The types of splitting modes are the same as those found for coplanar SpPGs: $\ell=0-9$ except $\ell=8$. The full results are given in the Supporting Information (Tables \spinsplitnocoplanar\, and \spinsplitnocoplanarTR). Here we restrict ourselves to comment on some differences with respect to the coplanar case:
	
\begin{table}[h]
	\caption{\label{tab:non-coplanar} Number of non-coplanar SpPGs for each of the non-relativistic spin splitting types.}
		\begin{tabular}{|c|c|c|c|c|c|c|c|c|c|c|c|c|}
	\hline
	Type&No splt.&$s$ (FM)&$p$&$d$&$f$&$g$&$h$&$i$&$j$&$k$&$\ell$&Total\\
	of splt.&(with $\overline{1}^{\prime}$)&$\ell=0$&$\ell=1$&$\ell=2$&$\ell=3$&$\ell=4$&$\ell=5$&$\ell=6$& $\ell=7$&$\ell=8$&$\ell=9$&\\
	\hline
	without TR&80&253&128&69&44&15&1&8&0&0&0&598\\
	\hline
	with TR&36&0&78&0&39&0&5&0&2&0&2&162\\
	\hline
\end{tabular}
\end{table}

\begin{itemize}
	\item In general, the electronic polarization has all three components nonzero ($B_i\neq0,i=u,v,w$) for even-parity and odd-parity splitting types. Depending on the particular SpPG, special polarization configurations may also occur.
	
	\item The $h$-wave can take place without the presence of TR in the point group. As indicated in Table \ref{tab:non-coplanar} there is one SpPG with this characteristic, the cubic non-coplanar group $^{2_{001}}m^{6_{001}}\overline{3}^{2_{100}}m$ (32.589). The functional form of this $h$-wave is somewhat different from that of the coplanar cases. If the components of ${\bf k}$ are referred to the $({\bf a},{\bf b},{\bf c})$ cubic unit cell vectors, we have
	\begin{equation}
		{\bf B}=Ck_1k_2k_3\left(k_1^2+k_2^2-2k_3^2,k_1^2+k_3^2-2k_2^2,0\right)
	\end{equation}
	where $C$ is a global constant, and the components of ${\bf B}$ are expressed in an orthonormal basis with (001) along the hexagonal spin-space direction. The polarization direction is therefore on the plane perpendicular to the hexagonal axis of the spin operations. The same ${\bf k}$ dependence may also occur for other non-coplanar groups with TR (Table \spinsplitnocoplanarTR\, in the Supporting Information). Figure \ref{fig:cubicwave} shows schematically the dependence of $B_u$ on the ${\bf k}$ direction around the $\Gamma$ point of the Brillouin zone, which can be compared with Figure \ref{fig:waves}(a). The function corresponds to the sum of two $h$-waves of the type found in coplanar structures, as $k_1^2+k_2^2-2k_3^2=(k_1^2-k_3^2)+(k_2^2-k_3^2)$.
	
	\begin{figure}[ht]
		\centering
		\includegraphics[width=0.4\textwidth]{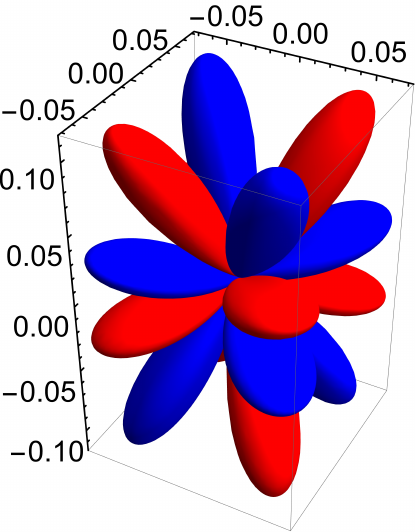}
		\caption{\label{fig:cubicwave}Schematic representation of $B_u$ near the center of the Brillouin zone for the $h$-wave of the cubic non-coplanar group $^{2_{001}}m^{6_{001}}\overline{3}^{2_{100}}m$ (32.589). Red (blue) curves indicate positive (negative) splittings. The components of the wave vector are in arbitrary units.}
	\end{figure}
	
	\item The presence of TR in the point group is required to raise the splitting type to the two highest $\ell$ values ($\ell=7,9$). These splittings occur in two hexagonal and two cubic noncentrosymmetric groups, respectively.
	
	Table \ref{tab:non-coplanarfunctions} shows the non-coplanar SpPGs with TR which permit the existence of $h$, $j$, and $\ell$-waves. In the orthonormal reference frame $(u,v,w)$ of spin space in which the spin operations are defined, ${\bf B}$ patterns near ${\bf k}=0$ have either three or two nonzero components.
	
		\begin{table}[h]
		\caption{\label{tab:non-coplanarfunctions} SpPGs with TR, as indicated by the $1^{\prime}$ suffix at the end of the SpPG labels, that allow the $h$, $j$, and $\ell$-wave spin splittings. In the expressions for ${\bf B}({\bf k})$ the components $B_i$ are referred to an orthonormal spin-space basis to which the spin operations are defined. For $\ell=7$, the components of ${\bf k}$ are given in an orthonormal setting with axes parallel to ${\bf a}$, ${\bf a}+2{\bf b}$, and ${\bf c}$. For $\ell=5$ and 9, an orthonormal setting with axes parallel to ${\bf a}$, ${\bf b}$, and ${\bf c}$ is used. The coefficients $C_i$ ($i=1,2,3$) are arbitrary constants.}
		\begin{tabular}{|c|c|c|c|c|c|c|}
			\hline
			Type of splt.&SpPGs&Spin splitting and polarizations\\
			\hline
			\multirow{6}{*}{$h$ ($\ell=5$)}&$^{1}4^{1}2^{1}2.1^{\prime}$ (12.130.1')&${\bf B}=k_1k_2k_3\left(k_1^2-k_2^2\right)\left(C_1,C_2,C_3\right)$\\
			&$^{1}4/^{2}m^{2}m^{2}m.1^{\prime}$ (15.199.1')&${\bf B}=k_1k_2k_3\left(k_1^2-k_2^2\right)\left(C_1,C_2,0\right)$\\
			&$^{2}m^{6}\overline{3}.1^{\prime}$ (29.546.1')&${\bf B}=C_1k_1k_2k_3\left(k_1^2-k_2^2,k_3^2-k_1^2,0\right)+$\\
			&&$+C_2k_1k_2k_3\left(k_3^2-k_2^2,k_2^2-k_1^2,0\right)$\\
			&$^{2_{100}}\overline{4}^{3_{001}}3^{2_{100}}m.1^{\prime}$ (31.555.1')&${\bf B}=C_1k_1k_2k_3\left(k_1^2+k_2^2-2k_3^2,k_2^2+k_3^2-2k_1^2,0\right)$\\
			&$^{2_{001}}m^{6_{001}}\overline{3}^{2_{100}}m.1^{\prime}$ (32.589.1')&${\bf B}=C_1k_1k_2k_3\left(k_1^2+k_2^2-2k_3^2,k_1^2+k_3^2-2k_2^2,0\right)$\\
			\hline
			\multirow{2}{*}{$j$ ($\ell=7$)}&$^{1}6^{1}2^{1}2.1^{\prime}$ (24.338.1')&${\bf B}=k_1k_2k_3\left(3k_1^2-k_2^2\right)\left(k_1^2-3k_2^2\right)\left(C_1,C_2,C_3\right)$\\
			&$^{1}6/^{2}m^{2}m^{2}m.1^{\prime}$ (27.453.1')&${\bf B}=k_1k_2k_3\left(3k_1^2-k_2^2\right)\left(k_1^2-3k_2^2\right)\left(C_1,C_2,0\right)$\\
			\hline
			\multirow{2}{*}{$\ell$ ($\ell=9$)}&$^{1}4^{1}3^{1}2.1^{\prime}$ (30.559.1')&${\bf B}=k_1k_2k_3\left(k_1^2-k_2^2\right)\left(k_2^2-k_3^2\right)\left(k_3^2-k_1^2\right)\left(C_1,C_2,C_3\right)$\\
			&$^{2}m^{2}\overline{3}^{2}m.1^{\prime}$ (32.574.1')&${\bf B}=k_1k_2k_3\left(k_1^2-k_2^2\right)\left(k_2^2-k_3^2\right)\left(k_3^2-k_1^2\right)\left(C_1,C_2,0\right)$\\
			\hline
		\end{tabular}
	\end{table}
	
\end{itemize}
\end{enumerate}
 \section{Conclusions}
\label{sec:conclusions}
In summary, we have carried out a systematic symmetry analysis of non-relativistic spin splitting in non-collinear magnetic structures using spin-group theory.

As a first step, a comprehensive enumeration and classification of non-equivalent SpPGs has been carried out, including those that contain the TR operation. This tabulation, which extends previous work, is freely available in the Bilbao Crystallographic Server with the name \emph{SPGENPOS}. 

The symmetry constraints on the spin splitting under each of these 1249 tabulated SpPGs have been determined applying the program \emph{STENSOR}, also in the Bilbao Crystallographic Server. This analysis shows that, in contrast with collinear systems, the presence of TR in the point group is not sufficient in general to forbid any spin splitting. Non-collinear systems, therefore, allow a much broader range of spin-splitting behaviors than collinear magnets, with spin-splitting waves ranging from $\ell=0$ to $\ell=9$, except for $\ell=8$. In particular, we demonstrate that $h$, $j$, and $\ell$-waves can occur in coplanar and non-coplanar structures with specific tetragonal, hexagonal, and cubic SpPGs. A preliminary survey of MAGNDATA has identified one coplanar tetragonal compound with $\ell=5$ non-relativistic spin-splitting. These results disprove arguments linking these waves to the requirement of rotational symmetries of order $\ell$, which would be incompatible with crystalline lattices. Moreover, we have found that odd-parity waves of low order (mainly $p$ and $f$, and in one non-coplanar case an $h$-wave) may arise in SpPGs without TR symmetry.

These results point to a relatively unexplored class of materials with non-relativistic spin splitting and could help broaden the range of magnetic states with potential relevance for applications or with fundamental interest. 		
\acknowledgments
	We thank Hana Schiff and Paul McClarty for helpful discussion about the tabulation of the spin point groups.

\onecolumngrid
\renewcommand{\thesection}{Appendix \arabic{section}}
\renewcommand{\thesubsection}{\arabic{section}.\arabic{subsection}}

\newpage

	\begin{center}{\bf Supporting Information of "Spin point group symmetry and classification of non-relativistic spin splitting in non-collinear magnetic structures: Identification of high-order spin splitting types ($\ell$=5,7, and 9)"}
		\end{center}

\tableofcontents

\clearpage

\addtocontents{toc}{\protect\setcounter{tocdepth}{3}}
\addtocontents{lot}{\protect\setcounter{lotdepth}{3}}
\renewcommand{\thetable}{S\arabic{table}}
\renewcommand{\thefigure}{S\arabic{figure}}
\renewcommand{\thesection}{S\arabic{section}}
\renewcommand{\thesubsection}{\thesection.\arabic{subsection}}
\renewcommand{\thesubsubsection}{\thesubsection.\arabic{subsubsection}}

\makeatletter
\def\p@subsection{}
\def\p@subsubsection{}
\makeatother

\setcounter{section}{0} 
\setcounter{table}{0} 
\setcounter{equation}{0}

\section{Tabulation of the spin point groups}
\label{sec:spgenpos}
In this section we explain the criteria used for the implementation of the tables of the spin point groups (SpPGs) in \emph{SPGENPOS}, taking as reference the first tabulation of the 598 non-coplanar SpPGs made by \citet{Litvin1977}, the identification of the 253 coplanar SpPGs \cite{Liu2022} and the 90 collinear groups \cite{Liu2022}. We have also determined the groups that result from these groups after the addition of time-reversal (TR) symmetry. 

Whereas Litvin assigned a number and a symbol to each of the non-coplanar groups, there is no numeration of the coplanar or collinear SpPGs. In \emph{SPGENPOS} we propose a systematic numeration of the SpPGs with and without TR, and a simplification of the symbols used by other authors. We also propose a \emph{standard} setting for each SpPG and give the $U$ and $R$ matrices of each symmetry operation in that setting. We have tried to keep as close as possible to the standard settings used in the crystallographic point groups.

In this section we first introduce the tool \emph{SPGENPOS} and explain how to use it to get the symmetry operations of the chosen SpPG and how is organized the output shown by the program. Afterwards, we explain the notation chosen for every type of SpPGs. First, we introduce in section \ref{sec:noncoplanar} the notation proposed for the non-coplanar groups, in section \ref{sec:coplanar} the  notation and setting chosen for the coplanar groups and in section \ref{sec:collinear} the notation and setting used in collinear groups. In section \ref{sec:elementary-non-trivial} we present some examples of SpPGs that cannot be reduced as the direct product of one of the 598 non-trivial groups tabulated by Litvin and a spin-only group, a type of SpPG that has not usually been considered so far. Finally we tabulate in section \ref{sec:groupswithTR} the SpPGs that result after the addition of TR to the collinear, coplanar and non-coplanar groups.

\begin{figure}[h]
	\caption{\label{fig:mainmenu}Partial view of the main menu of \emph{SPGENPOS}. It shows the list of the 32 crystallographic point groups in the usual order and six boxes below the table that correspond to the six possible types of SpPGs to be chosen by the user.}
	\begin{center}
		\includegraphics[scale=0.5]{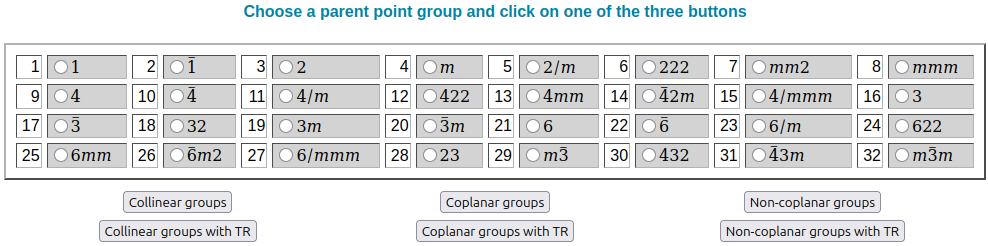}
	\end{center}
\end{figure}

 In \emph{SPGENPOS} the SpPG with symmetry operations $\{U||R\}$ is selected in two steps: in the first step the user chooses one of the 32 crystallographic point groups which corresponds to the point group formed by the $R$ operations in the orbital space. This group is often called \emph{parent point group} of the SpPG. Then, the user chooses the type of SpPG among the types of groups offered by the program: collinear, coplanar, non-coplanar or one of these types of groups combined with TR. Fig. \ref{fig:mainmenu} shows a partial view of the main page of \emph{SPGENPOS}, where the crystallographic point groups are ordered in the usual way. It is important to stress that, although there is no standard numeration of the 32 crystallographic point groups, they are usually ordered in the specific way shown in the figure. This order is implicitly assumed, for instance, in the standard numeration of the 230 space groups, and in the standard numeration of the magnetic point groups. We have included the number of the point group together with its symbol for reference. This numeration has been assumed in all the programs and databases of the Bilbao Crystallographic Server (see for instance the tool \href{https://cryst.ehu.es/rep/point.html}{POINT}). Once a parent point group has been chosen, clicking on one of the boxes below the table of point groups in Fig. \ref{fig:mainmenu}, the program shows in an intermediate page the list of SpPGs that correspond to the parent group and SpPG type chosen. As the number and notation used in \emph{SPGENPOS} depends on the type of group (non-coplanar, coplanar or collinear), in the next sections we describe separately these three cases.
 
\subsection{Non-coplanar groups}
\label{sec:noncoplanar}
The first complete identification of the 598 non-coplanar groups with a spin-only group equal to the identity (non trivial groups) was done by Litvin \cite{Litvin1977}. The groups are ordered first according to the point group formed by the operations $R$ in the orbital space (parent point group of the SpPG). Among the spin groups that share the same parent point group, they are ordered according to the point group formed by the subgroup of operations $\{1||R\}$. Finally, among the SpPGs that share both point groups, they are ordered according to the point group formed by the $U$ operations in spin space. In this way, to each different SpPG Litvin assigned a number from 1 to 598.

In \emph{SPGENPOS} we have assigned to each non-coplanar SpPG a number with the format $i.j$ to make more clear the classification of the SpPGs. The second integer $1\le j\le598$ is the integer assigned by Litvin in his original work \cite{Litvin1977}. The first integer $1\le i\le32$ is the number of the parent point group.  In general, if the numbers of two different SpPGs are $i.j$ and $i'.j'$, the relation $i>i'$ implies $j>j'$. However, in Litvin's classification, some pairs of point groups were interchanged with respect to the usual order shown in Fig. \ref{fig:mainmenu} as, for example, the point groups $222$  and $mm2$ (point group numbers 6 and 7 in Fig. \ref{fig:mainmenu}, respectively). Therefore, in our notation, the SpPGs with numbers $6.j$ and $7.j'$ fulfill $j>j'$. The other pairs of swapped point groups in Litvin's list with respect to the order assumed in \emph{SPGENPOS} are point groups 21 and 22, 23 and 24 and 30 and 31.

\begin{figure}[h]
	\caption{\label{fig:noncop622}List of non-coplanar SpPGs with parent point group 622 (N. 24).}
	\begin{center}
		\includegraphics[scale=0.5]{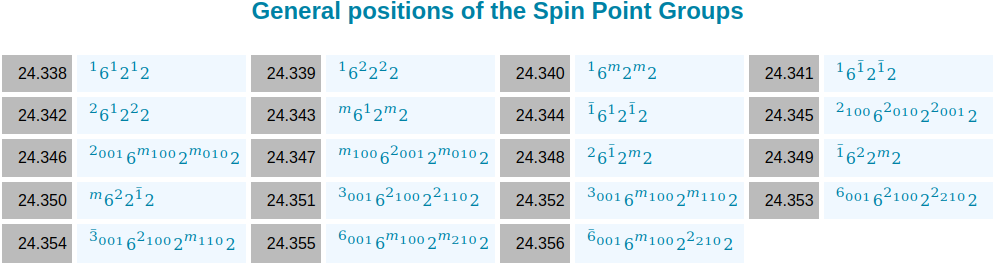}
	\end{center}
\end{figure}

Once the user has chosen one of the parent point groups, clicking on the box \emph{Non-coplanar groups}, the program shows the list of non-coplanar groups that correspond to the given parent group. Fig. \ref{fig:noncop622} gives the particular case for the parent point group 622 (N. 24). There are 19 different SpPGs denoted in \emph{SPGENPOS} as $24.338-24.356$. Together with the number, the program gives a symbol for each SpPG. These symbols are based on the list of symbols proposed by Litvin \cite{Litvin1977} but they have been slightly modified in some cases. To every symmetry operation $R$ of the orbital space included in the symbol, the symbol of the $U$ operation is attached to the symbol of $R$ as a pre-superscript $^{U}R$. As in the original notation proposed by Litvin \cite{Litvin1977}, when the set of $U$ operations form a monoclinic group, the $2$ (2-fold axis) or $m$ (mirror plane) symbols have no subscript (as in the SpPG group 24.339 in Fig. \ref{fig:noncop622}, for example). This choice reinforces the idea that the spin and orbital spaces are decoupled and, then, the direction of the 2-fold axis (or the direction perpendicular to the plane) is arbitrarily oriented with respect to the directions in the orbital space. In the rest of crystal systems (excluded the triclinic case), we have changed the subindex attached to the symmetry operation in spin space that denotes the direction kept invariant(reversed) after the application of the proper(improper) operation in spin space. Instead of the $x,y,z$ symbols used in Litvin's original paper (complemented by subscripts 1 and 2 in hexagonal point groups), we have used three integers $U,V,W$ that represent the direction $[UVW]$ in a standard orthogonal basis (orthorhombic, tetragonal and cubic groups) or in a hexagonal basis (trigonal and hexagonal groups). For instance, the group number 355 with symbol $^{6_{z}}6^{m_{x}}2^{m_{1}}2$ in Litvin's list has number 24.355 and symbol $^{6_{001}}6^{m_{100}}2^{m_{210}}2$ in \emph{SPGENPOS} (see Fig. \ref{fig:noncop622}).

Another significant change with respect to the list of symbols in Litvin's classification affects the SpPGs with parent group $m\overline{3}$ (N. 29) or $m\overline{3}m$ (N. 32). In Litvin's work, the notation of these parent groups was changed to $2/m\overline{3}$ and $4/m\overline{3}2/m$, respectively, and the corresponding $U$ operations were attached to every orbital $R$ operation that is included in the new symbol. However, one of the mirror planes $m$ and one of the roto-inversion axis $\overline{3}$ in group N. 29, or these two operations together with the second mirror plane $m$ in group N. 32, are a complete set of generators of the parent point group. Therefore, the two operations $\{U(m_{001})|m_{001}\}$ and $\{U(\overline{3}_{111})|\overline{3}_{111}\}$ in a SpPG with parent point group number 29 or these two operations together with $\{U(m_{110})|m_{110}\}$ in a SpPG with parent point group number 32 form a set of generators of the SpPG. The addition of the operation $\{U(2_{001})|2_{001}\}$ in the first case and the operations $\{U(4_{001})|4_{001}\}$ and $\{U(2_{110})|2_{110}\}$ in the second case is redundant. We have removed these extra operations from the symbol for two reasons: on the one hand, the standard symbol of the orbital part used in other contexts is maintained and, on the other hand, the symbol of the SpPG is significantly simplified, without introducing any ambiguity.

\begin{figure}[h]
	\caption{\label{fig:symop}Partial list of symmetry operations of the SpPG $^{m_{100}}6^{2_{001}}2^{m_{010}}2$ (N. 24.347). The first column represents a sequential number of the symmetry operations. In the second column the operations are given in the $(x,y,z,u,v,w)$ format. The third and fourth columns give the matrices $U$ and $R$ and the last column shows the generalized Seitz notation $\{U||R\}$ of the operation. The anti-unitary operations (those with $\det(U)=-1$) are colored in red.}
	\begin{center}
		\includegraphics[scale=0.8]{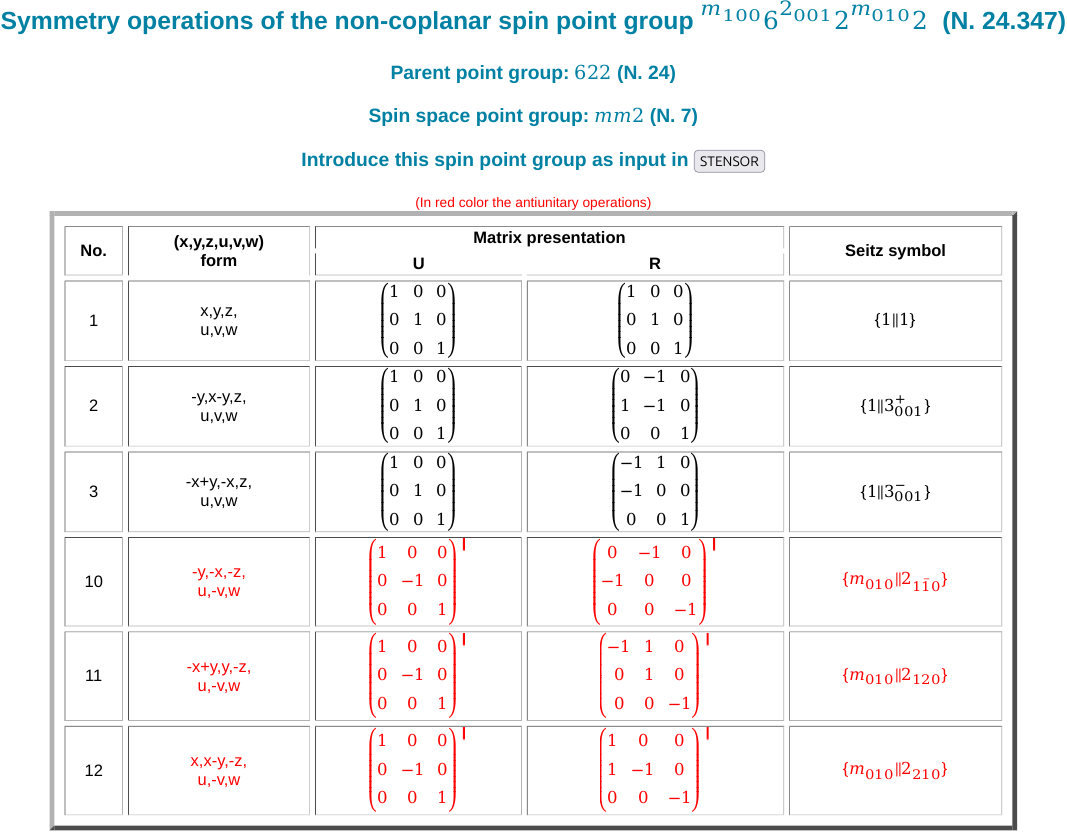}
	\end{center}
\end{figure}

In the last step, clicking on one of the symbols listed in the intermediate menu, the program gives the list of symmetry operations in a specific setting. In the orbital space we have chosen an hexagonal basis ($a=b$, $\alpha=\beta=90^{\circ}$ and $\gamma=120^{\circ}$) for trigonal and hexagonal point groups and an orthogonal basis ($\alpha=\beta=\gamma=90^{\circ}$) for the rest of crystal systems to write the matrices of the $R$ operations. We have used the same convention also in the spin space, an hexagonal basis ($a_s=b_s$, $\alpha_s=\beta_s=90^{\circ}$ and $\gamma_s=120^{\circ}$) for trigonal and hexagonal point groups and an orthogonal basis ($\alpha_s=\beta_s=\gamma_s=90^{\circ}$) for the rest of crystal systems to write the matrices of the operations $U$. Note that in some SpPGs, the basis chosen in one space can be orthogonal and hexagonal in the other space. We have chosen the ${\bf b}$ axis parallel(perpendicular) to the 2-fold axis(mirror plane) in the orbital space when the point group is monoclinic, but the \emph{special} direction has been chosen parallel to the ${\bf c}_s$ direction in the spin space. For instance, in the monoclinic group SpPG $^{m}2$ (N. 3.8), the 2-fold axis in the orbital space is parallel to the basis vector ${\bf b}$, but the mirror plane in the spin space is perpendicular to ${\bf c}_s$. Fig. \ref{fig:symop} shows some symmetry operations (the full set has been cut to keep a reasonable size of the figure) of the SpPG $^{m_{100}}6^{2_{001}}2^{m_{010}}2$ (N. 24.347). The output of the program gives the whole list of symmetry operations in different formats: the $(x,y,z,u,v,w)$ format, the explicit $3\times3$ matrices $U$ and $R$, and the generalized Seitz symbol $\{U||R\}$. The anti-unitary operations (those with $\det(U)=-1$) are colored in red. The matrices of the anti-unitary operations have also a \emph{prime} symbol attached as superscript as in other databases in the \emph{Magnetic Symmetry and Applications} section of the Bilbao Crystallographic Server. Just above the table, the program gives the parent point group formed by the $R$ operations and the spin space point group formed by the $U$ operations.

The output also includes a direct link to \emph{STENSOR}, also located at the Bilbao Crystallographic Server. This program calculates the symmetry reduction of any tensor (with rank $r\le8$) under the chosen SpPG (see \citet{etxebarria2025} and \citet{Elcoro2026} for information about \emph{STENSOR}).

\subsection{Coplanar groups}
\label{sec:coplanar}
All non-coplanar SpPGs discussed in the previous section whose spin point group formed by the $U$ operations is non-cubic are compatible with a coplanar distribution of spins. The trigonal, tetragonal and hexagonal spin point groups are compatible with a coplanar distribution where the spins are contained in the plane perpendicular to the 3-, 4-, 6-fold axis or to these axes multiplied by the inversion (roto-inversion axes). In triclinic, monoclinic and orthorhombic spin point groups the plane of spins compatible with the symmetry operations in the spin space can take different orientations. The systematic search of all possible SpPGs of coplanar distributions of spins can be performed taking all the 598 groups tabulated by Litvin and implemented in \emph{SPGENPOS} following the criteria stated in section \ref{sec:noncoplanar} and adding to these groups the symmetry operation $\{m_{uvw}||1\}$ in all the possible orientations compatible with the operations in the chosen non-coplanar group. Afterwards, we need to consider possible equivalences between the newly formed sets of operations to avoid duplicities in the tabulation of the groups. Two sets of operations $\{U_i||R_i\}$ and $\{U_j'||R_j'\}$ are equivalent, and thus form the same group-type, if there exists a transformation matrix $\{P^U||P^R\}$, $P^U$ and $P^R$ being non-singular $3\times3$ matrices, such that we can establish a 1:1 relation between the operations in both sets in this way,
\begin{equation}
	\label{eq:equivalence}
\{U_j'||R_j'\} = \{(P^U)^{-1}U_iP^U||(P^R)^{-1}R_iP^R\}
\end{equation}
It is said that these two subsets correspond to the same group-type (SpPG-type in this case) and the transformation matrix $\{P^U||P^R\}$ represents a change of basis (or change of setting) in the spin and orbital subspaces. Among all the subsets that correspond to the same SpPG-type only one of them must be considered in the tabulation of the coplanar SpPGs. The determination of the 253 resulting coplanar SpPGs and the assignation of a symbol to each group was recently performed by \citet{Liu2022} (see tables S1-S11 in the Supplementary Material of this reference).

In \emph{SPGENPOS} we have adopted several conventions that do not coincide, in general, with the conventions in \citet{Liu2022} and we have also added a number to each coplanar SpPG.

In a coplanar SpPG, to every operation $R$ of the orbital subspace correspond two different operations in the spin subspace, i.e., there are two operations $\{U||R\}$ and $\{mU||R\}$ in the SpPG, being $m$ the mirror plane in the spin subspace parallel to the coplanar distribution of spins. One of the two operations $U$ or $mU$ is a proper operation. Following \citet{Liu2022} in the construction of the symbol of the coplanar group, between the two possible choices to denote the spin part of every symmetry operation that appears in the symbol of the SpPG we have taken the proper operation in all the cases. If we consider the index 2 subgroup of the coplanar SpPG that contains only the operations whose spin part is a proper operation, these operations form a non-coplanar group that must be equivalent to one of the 598 non-coplanar tabulated groups of section \ref{sec:noncoplanar}. Therefore, each coplanar group can be expressed as the direct product of a non-coplanar group whose operations have a proper spin part and a SpPG that contains two elements: the identity and $\{m_{uvw}||1\}$.

It is important to stress at this point that to every coplanar group it corresponds only one non-coplanar group as reference, but different coplanar groups can have the same non-coplanar group as reference. For example, the non-coplanar SpPG $^{2}\overline{1}$ (N. 2.3 in \emph{SPGENPOS}) is the reference group of two different coplanar SpPGs, depending on the relative orientation of the 2-fold axis and the plane of spins, due to the fact that a 2-fold axis keeps invariant the plane perpendicular to the axis and also any plane that contains the axis. In \citet{Liu2022}, these two different coplanar groups are denoted as $^{2_z}\overline{1}^{m_z}1$ and $^{2_z}\overline{1}^{m_x}1$. In the first case the 2-fold axis is perpendicular to the plane of spins and in the second case it is parallel. In this notation, the non-trivial part of the group is common to both groups ($^{2_z}\overline{1}$) and they differ in the orientation of the plane of the trivial group: $^{m_z}1$ and $^{m_x}1$, respectively. In \emph{SPGENPOS} we have taken a different criterion. In all the 253 coplanar groups we have chosen a setting in which the mirror plane of the trivial spin-only group is perpendicular to the $w$ direction (third direction in the assumed orthogonal or hexagonal basis in spin space). As a consequence, all the 253 coplanar groups share the same trivial spin-only subgroup in the same setting and, then, the operations of the non-trivial subgroups of two coplanar groups that have the same reference non-coplanar group are different. In our previous example, the coplanar groups $^{2_z}\overline{1}^{m_z}1$ and $^{2_z}\overline{1}^{m_x}1$ in \citet{Liu2022} are denoted in \emph{SPGENPOS} as $^{2_{001}}\overline{1}.P$ and $^{2_{100}}\overline{1}.P$, respectively. Note that we have substituted the trivial part of the notation ($^{m_{001}}1$), common to all coplanar groups, by $.P$. We have also defined a number for each coplanar SpPG using the format $i.j.k.P$. The $i.j$ pair of integers denote the non-coplanar reference group and $1\le k\le253$ is a correlative number to index all the 253 coplanar groups. As different (up to three) coplanar SpPGs can have the same non-coplanar SpPG as reference, the coplanar groups that share the same reference group have correlative $k$ indices. Finally, the last part of the symbol, $.P$, indicates that it is the number of a coplanar group.

The way to get the symmetry operations of the coplanar groups in \emph{SPGENPOS} follows the same two steps as in the non-coplanar case. First the user chooses the parent point group and, clicking on the button \emph{Coplanar groups} shown in Fig. \ref{fig:mainmenu}, the program goes to an intermediate page similar to the one shown in Fig. \ref{fig:noncop622}. Finally, clicking on the symbol of a coplanar group, the program lists all the symmetry operations of the SpPG. For coplanar groups, the program lists all the operations of the group. The first half of operations are  those with proper $U$ and the second half corresponds to the operations with improper $U$, colored in red.
\subsection{Collinear groups}
\label{sec:collinear}
All non-coplanar SpPGs discussed in section \ref{sec:noncoplanar} whose spin point group formed by the $U$ operations is non-cubic are also compatible with a collinear distribution of spins. In this type of spin distributions, all the spins are parallel to a given direction of the spin subspace. Such a distribution of spins remains invariant under any rotation of any infinitesimal angle with respect to an axis parallel to the spins (this rotation is denoted as an $\infty$ axis) and under any mirror plane that contains the axis. This trivial spin-only SpPG is denoted as $^{\infty_{\bf n}m}1$, where ${\bf n}$ identifies the direction of the axis.

In principle, the search of all different collinear SpPGs can be done following the same procedure used in the determination of the coplanar groups in section \ref{sec:coplanar}, resulting in the 90 non-collinear groups isomorphic to the 90 non-gray MPGs.
Following an equivalent convention as the one used for coplanar groups, in all the collinear groups we have taken the direction of spins ${\bf n}$ along $w$ (the third direction in the spin space). 

Every collinear group has thus a reference non-coplanar SpPG and, contrary to what happens in coplanar SpPGs, no different collinear groups share the same reference non-coplanar group. As in the coplanar case, the symbol of a collinear SpPG is composed by the symbol of its reference SpPG followed by $.L$. Note that, as we have decided to choose the infinite axis parallel to $w$ in all the cases, it is not necessary to make it explicit in the notation of the group. The number of the collinear groups has the format $i.j.k.L$, being $i.j$ the number of the reference non-coplanar group, the third integer is a correlative number $1\le k\le90$, and the last term $.L$ indicates that it is a collinear group.

The symmetry operations can be obtained following the two-step procedure explained above for non-coplanar and coplanar groups. In this case, however, the list of operations is displayed in two subsets: first it is included the list of operations of the non-trivial group and, in a second subset, the operations of the trivial group are displayed in parametric form for the $U$ components: a first matrix that represents a rotation of angle $\varphi$ around the $w$ axis, and a mirror plane that contains the $w$ axis and whose normal is rotated an angle $\varphi$ with respect to the $u$ direction.
 \subsection{Examples of an SpPGs not expressible as direct product of non-trivial and spin-only subgroups}
\label{sec:elementary-non-trivial}
The SpPGs of most magnetic structures satisfy the decomposition given by equation (\ref{equationationdirectproduct}) in the main text, with \Pnt\, being a so-called non-trivial group not having spin-only operations except the identity, and \Pso\, being a group formed by all the spin-only operations.  But, contrary to the commonly accepted assumption, there are cases where this decomposition is not possible and a more general expression must be used, 
\begin{equation}
	\label{eq:directproductSM1}
	P_{\textrm{\footnotesize{S}}}=P_{\textrm{\footnotesize{E}}}\times P_{\textrm{\footnotesize{SO}}}
\end{equation}

where \Pe\, may contain spin-only operations different from the identity. 

Consider a non-coplanar spin space group with the operation $\{4_{001}^{+}||2_{010}|0,0,1/4\}$ that, together with the three translations $\{1||1|1,0,0\}$, $\{1||1|0,1,0\}$ and $\{1||1|0,0,1\}$ generates the spin space group. The 2-fold and 4-fold axes are along the $(0,y,0)$ and $(0,0,w)$ directions in real space and spin space, respectively, following the conventions introduced in section \ref{sec:spgenpos}. The coset representatives  of the full space group $G_S$ with respect to the subgroup of lattice translations are then:
\begin{equation}
	\{1||1|0,0,0\},\{4_{001}^{+}||2_{010}|0,0,1/4\},\{2_{001}||1|0,0,1/2\},\{4_{001}^{-}||2_{010}|0,0,3/4\}
\end{equation}
The corresponding SpPG is then formed by the four operations:
\begin{equation}
	P_{\textrm{\footnotesize{S}}}=\{\{1||1\},\{4_{001}^{+}||2_{010}\},\{2_{001}||1\},\{4_{001}^{-}||2_{010}\}\}
\end{equation}
It is straightforward to verify that whereas G$_{\textrm{\footnotesize{S}}}$=G$_{\textrm{\footnotesize{NT}}}$ $\times$ G$_{\textrm{\footnotesize{SO}}}$, with G$_{\textrm{\footnotesize{NT}}}$=G$_{\textrm{\footnotesize{S}}}$, and the spin-only subgroup being reduced to the identity, G$_{\textrm{\footnotesize{SO}}}=\{1||1|000\}$, it is impossible to write an analogous equation for \Ps\, since the presence of the spin-only operation $\{2_{001}||1\}$ prevents its classification as non-trivial in the sense established by \citet{Litvin1977}. This contrasts with the spin space group case, where the spin-translation group operation $\{2_{001}||1|0,0,1/2\}$ is not an spin-only operation, and it therefore belongs to the non-trivial G$_{\textrm{\footnotesize{NT}}}$. A preliminary review of the spin space groups obtained using the program \emph{FindSpinGroup} \cite{Yu2026} of the structures listed in MAGNDATA \cite{Gallego2016} indicates that about 0.5\% of the spin space groups in real materials have a SpPG of this type, and therefore cannot be related by mean of a direct product with one of the 598 non-trivial SpPGs.
Table \ref{tab:rarecases} lists some examples obtained with \emph{FindSpinGroup} \cite{Yu2026} of structures of MAGNDATA that do not allow such a direct product. 

\begin{center}\begin{table}
\caption{\label{tab:rarecases}Some examples of experimental structures in MAGNDATA whose SpPG, derived from the spin space group obtained by \emph{FindSpinGroup} (Yu et al. (2026)),  cannot be expressed as the direct product of a non-trivial point group and a spin-only point group. The first column indicates the MAGNDATA code of the compound, the second column shows the type of spin distribution (collinear, coplanar, non-coplanar), the third column gives the elementary SpPG, \Pe, and the last column the generators of the spin-only subgroup of \Pe\, (see section \ref{sec:groupswithTR}).}
	\begin{tabular}{|c|c|c|c|}
	\hline
	code&type&\Pe&P$_{E}^{SO}$\\
	\hline
	1.203&cop&$^{4_{001}}m\,^{1}m\,^{4_{001}}m$&$\{2_{001}||1\}$\\
	1.528&&&\\
	\hline
	1.380&cop&$^{2_{010}}4/\,^{2_{110}}m\, ^{2_{010}}m\,^{1}m$&$\{2_{001}||1\}$\\
	1.381&&&\\
	1.382&&&\\
	1.383&&&\\
	1.385&&&\\
	1.386&&&\\
	1.387&&&\\
	1.844&&&\\
	\hline
	1.839&cop&$^{4_{001}}4/\,^{2_{110}}m\,^{2_{100}}m\,^{2_{110}}m$&$\{2_{001}||1\}$\\
	\hline
	1.0.32&cop&$^{6_{001}}6/\,^{2_{100}}m\,^{1}m\,^{6_{001}}m$&$\{3_{001}^{+}||1\}$\\
	\hline
	2.22&cop&$^{2_{100}}4/\,^{2_{110}}m$&$\{2_{001}||1\}$\\
	\hline
	2.86&cop&$^{4_{001}}4/\,^{2_{110}}m$&$\{2_{001}||1\}$\\
	\hline	
\end{tabular}
\end{table}\end{center}

An example of SpPG of a real structure that cannot be expressed as P$_{\textrm{\footnotesize{NT}}}$ $\times$ P$_{\textrm{\footnotesize{SO}}}$ is that of LaMnAu$_5$ (N. 1.839 in MAGNDATA), which is discussed in section \ref{sec:classification} of the main text. The MPG of this structure is $4mm.1’$, generated by the operations $\{4_{001}^{+}||4_{001}^{+}\}$, $\{2_{100}||m_{100}\}$ and $\{-1||1\}$, while its SpPG is augmented by the operation $\{2_{110}||m_{001}\}$. The full group has 64 operations and can be written as $	P_{\textrm{\footnotesize{S}}}=P_{\textrm{\footnotesize{E}}}\times\,^{m_{001}}1$ where $^{m_{001}}1$ is the intrinsic coplanar group. However, $P_{\textrm{\footnotesize{E}}}$ is not a non-trivial point group. It has 32 operations (generated by $\{4_{001}^{+}||4_{001}^{+}\}$, $\{2_{100}||m_{100}\}$, and $\{2_{110}||m_{001}\}$), and includes the spin-only group operation $\{2_{001}||1\}$.
\subsection{SpPS of the spin space groups with time-reversal symmetry}
\label{sec:groupswithTR}
In general, the list of possible SpPGs relevant for the analysis of physical properties must be extended to account for cases in which the group $\mathcal{T}_A$ of lattice translations of the idealized atomic positions (or the electronic density) is different from the group of translations $\mathcal{T}_S$ of the spins associated to the atoms, i.e., there are translations in $\mathcal{T}_A$ that transform the structure into itself when the spins are not considered, but atoms separated by ${\bf t}\in\mathcal{T}_A$ with ${\bf t}\notin\mathcal{T}_S$ have their spins pointing along different directions. If the spins of all pairs of atoms separated by ${\bf t}$ are related by a point group operation $U$, then $\{U||1|{\bf t}\}$ is a symmetry operation that belongs to the spin space group of the magnetic structure. In all cases, it is possible to express the SpPG as the direct product of two SpPGs (see equation (\ref{eq:directproductSM1})),
\begin{equation}
	\label{eq:directproductSM}
	P_{\textrm{\footnotesize{S}}}=P_{\textrm{\footnotesize{E}}}\times P_{\textrm{\footnotesize{SO}}}=P_{\textrm{\footnotesize{E}}}\times P_{\textrm{\footnotesize{SOintr}}}\times P_{\textrm{\footnotesize{SOG}}}
\end{equation}
\begin{itemize}
\item \Ps\, is the SpPG of the spin space group formed by all the operations $\{U||R\}$. 
\item \Pso\, is a group formed by only $\{U||1\}$ (spin-only) operations. Note, however, that, in general, not all operations in \Ps\, of the form $\{U||1\}$ belong to \Pso. Recall the examples of Sec. \ref{sec:elementary-non-trivial}. This group can be expressed in turn as the direct product of two SpPGs, $P_{\textrm{\footnotesize{SO}}}=P_{\textrm{\footnotesize{SOintr}}}\times P_{\textrm{\footnotesize{SOG}}}$. 
\item $P_{\textrm{\footnotesize{SOintr}}}$ is a intrinsic SpPG and it is limited to $^{1}1$, $^{m}1$ and $^{\infty m}1$ for non-coplanar, coplanar and collinear structures, respectively. The spin space group contains thus the operations $\{U||1|{\bf T}\}$, where $U$ are the spin space operations of the corresponding trivial group and the translation ${\bf T}$ belongs to $\mathcal{T}_S$.
\item \Psog\, is formed by a subset of the point-group operations of the form $\{U||1\}$ associated with spin-space operations $\{U||1|{\bf t}\}$ of the spin translation group. In general, not every operation of the form $\{U||1\}$ is included in \Psog\, (see the examples of Sec. \ref{sec:elementary-non-trivial}).
\item \Pe\, is a SpPG that we call \emph{elementary}. We define the elementary SpPGs as those \Pe\, groups that cannot be written as,
\begin{equation}
	\label{eq:elementary}
	P_{\textrm{\footnotesize{E}}}=P_{\textrm{\footnotesize{E}}}^{'}\times P_{\textrm{\footnotesize{SO}}}
\end{equation}
being $P_{\textrm{\footnotesize{E}}}^{'}$ a SpPG and \Pso\, a spin-only SpPG different from the trivial one, i.e., different from the group that contains only the identity. Note that the expression (\ref{eq:directproductSM}) generalizes the expression (8) in \citet{etxebarria2025}, where \Pnt\, was used instead of \Pe\, being \Pnt\, one of the 598 non-trivial groups in Litvin's tabulation \cite{Litvin1977}. The non-trivial groups are in fact a subset of the elementary groups defined by equation (\ref{eq:elementary}), but not all the elementary groups are non-trivial. There are \Pe\, groups that contain spin-only operations and do not belong thus to the subset of non-trivial groups. In section \ref{sec:elementary-non-trivial} we have presented some examples of elementary SpPGs that contain spin-only operations.

Another consequence of the existence of elementary groups that are not non-trivial groups is that the SpPG \Pso\, in equation (\ref{eq:directproductSM}) does not contain all the spin-only operations of \Ps. If we define 
the normal subgroup $P_{\textrm{\footnotesize{E}}}^{\textrm{\footnotesize{SO}}}$ of \Pe, i.e., $P_{\textrm{\footnotesize{E}}}^{\textrm{\footnotesize{SO}}}\trianglelefteq P_{\textrm{\footnotesize{E}}}$ as the set of spin-only operations in the elementary SpPG \Pe, the \emph{full spin-only} group of the spin space group is,
\begin{equation}
	P_{\textrm{\footnotesize{FSO}}}=P_{\textrm{\footnotesize{E}}}^{\textrm{\footnotesize{SO}}}\times P_{\textrm{\footnotesize{SO}}}=P_{\textrm{\footnotesize{SOintr}}}\times P_{\textrm{\footnotesize{E}}}^{\textrm{\footnotesize{SO}}}\times P_{\textrm{\footnotesize{SOG}}}
\end{equation}
If the elementary SpPG is a non-trivial group, obviously $P_{\textrm{\footnotesize{FSO}}}=P_{\textrm{\footnotesize{SO}}}$.
\end{itemize}
When the set of atomic and spin translations coincide, \Psog\, contains only the identity, the elementary SpPG is a non-trivial group and thus \Ps\, is one of the 941 SpPGs of sections \ref{sec:noncoplanar}, \ref{sec:coplanar} and \ref{sec:collinear}. However, when $\mathcal{T}_A$ and $\mathcal{T}_S$ are different (but $\mathcal{T}_S$ is a subgroup of $\mathcal{T}_A$), the order of $P_{\textrm{\footnotesize{E}}}^{\textrm{\footnotesize{SO}}}\times P_{\textrm{\footnotesize{SOG}}}$ is the subgroup index $\mathcal{T}_A:\mathcal{T}_S$. In principle, this index is not bounded and, thus, the point group can be of any finite (or even infinite) order. This group can be any crystallographic point group, one of the icosahedral point groups or a cyclic or dihedral group that contains a $n$-fold axis (or $\overline{n}$ roto-inversion operation) of arbitrary $n$. In coplanar structures, the cubic and icosahedral point groups are not compatible and, if $P_{\textrm{\footnotesize{E}}}^{\textrm{\footnotesize{SO}}}\times P_{\textrm{\footnotesize{SOG}}}$ is a cyclic or dihedral group that contains a $n$-fold axis (or roto-inversion operation $\overline{n}$) with $n>2$, the axis must be perpendicular to the plane of spins. In collinear groups \Pe\, can always be chosen to be one of the non-trivial groups. Moreover, in collinear structures there are only two possible \Psog\, groups that result in different SpPGs for each \Pnt\, in equation (\ref{eq:directproductSM}) with $P_{\textrm{\footnotesize{E}}}=P_{\textrm{\footnotesize{NT}}}$: the trivial group 1 and the group that contains the identity and TR, point group $\overline{1}$.

As stressed above, several research groups \cite{Chen2024,Jiang2024,Xiao2024} have recently tabulated thousands of spin space groups and their corresponding SpPGs. These tabulations include cyclic and dihedral groups with axes (or roto-inversion) of order up to 48. On the other hand, \citet{Chen2024} have identified the spin space group of more than 2000 experimentally reported commensurate magnetic structures taken from the MAGNDATA database \cite{Gallego2016}. The results indicate that:
\begin{itemize}
	\item Except in rare cases (see section \ref{sec:elementary-non-trivial}), the elementary group in the direct product (\ref{eq:directproductSM}) is a non-trivial group.
	\item  About 57\% of the identified spin space groups have a SpPG whose \Psog\, group in the decomposition given by equation (\ref{eq:directproductSM}) is the group that contains only the identity.
	\item About 42\% of the structures have $P_{\textrm{\footnotesize{SOG}}}:\,^{\overline{1}}1$. These cases correspond to the materials whose magnetic space group is of type IV and contain anti-translations. When the relativistic effects cannot be disregarded, the operations $\{-1||1|{\bf t}\}$ still remain relevant, since they are symmetry operations of the magnetic space group.
	\item The remaining $\approx1$\% of commensurate magnetic materials in MAGNDATA are structures whose \Psog\, group is different from point groups $^{1}1$ and $^{\overline{1}}1$. These operations are not relevant when relativistic effects are important and they do not remain as symmetry operations of the magnetic space group.
\end{itemize}
The above results indicate that, apart from the cases whose \Psog\, group contains only the identity, the \Psog\, group that contains the identity and TR is by far the most relevant one in the analysis of the physical properties of magnetic materials. We have decided thus to add to SPGENPOS the SpPGs that contain the TR in the three types of groups: non-coplanar, coplanar and collinear groups. We have not considered groups \Pe\, different from the 598 non-trivial groups.

\begin{enumerate}
\item {\bf Non-coplanar SpPGs with time-reversal symmetry}

The addition of TR to the 598 non-coplanar groups gives sets of symmetry operations that are equivalent. Two sets of operations $\{U||R\}$ that share the same point group type in the orbital space but whose spin point group types are different cannot be equivalent. They must be assigned to two different entries in Litvin's list. However, if the two spin groups belong to the same Laue class, they become into the same point group type after the addition of TR. Therefore, after the addition of TR to the 598 non-coplanar groups, there will be sets of equivalent groups among them. Grouped into subsets of groups that share the same point group types both in orbital and spin subspaces, we have performed a systematic search of equivalences between these newly formed groups using equation (\ref{eq:equivalence}). We have identified 162 different groups. Table \ref{tab:equinoncoplanar} summarizes the results of the determination of the non-coplanar groups with TR. The first column gives the number assigned to each non-equivalent non-coplanar SpPG with TR. In the second and third columns, for each non-equivalent group with TR, the boxes contain the number and symbols of the non-coplanar SpPGs that, after the addition of TR, become equivalent. Note that the number in the first column is just the number of the first non-coplanar SpPG in the second column followed by $.1'$. The first non-coplanar SpPG has been chosen as the reference group to construct the SpPG with TR. The last column contains one box for each non-equivalent group and includes the following information: the first row shows the symbol of the non-coplanar group with TR, and it has been constructed just attaching $.1'$ to the symbol of the non-coplanar group chosen as reference. The next rows indicate the transformation matrix that relates the symmetry operations of the corresponding SpPG with TR to the setting of the first group in the box (reference group). The transformation matrix $\{P^R||P^U\}$ is related to the change of basis necessary to transform a set of symmetry operations in matrix form into the set of matrices in the standard setting chosen,
\begin{equation}
	\label{eq:changebasis}
\begin{array}{rcl}
	 {\bf a}',{\bf b}',{\bf c}'&=&P_{11}^R{\bf a}+P_{21}^R{\bf b}+P_{31}^R{\bf c},P_{12}^R{\bf a}+P_{22}^R{\bf b}+P_{32}^R{\bf c},P_{13}^R{\bf a}+P_{23}^R{\bf b}+P_{33}^R{\bf c}\\
	 {\bf a}_s',{\bf b}_s',{\bf c}_s'&=&P_{11}^U{\bf a}_s+P_{21}^U{\bf b}_s+P_{31}^U{\bf c}_s,P_{12}^U{\bf a}_s+P_{22}^U{\bf b}_s+P_{32}^U{\bf c}_s,P_{13}^U{\bf a}_s+P_{23}^U{\bf b}_s+P_{33}^U{\bf c}_s	 
\end{array}\end{equation}
The table includes only those changes of basis different from the identity. In most cases, the two sets of symmetry operations require no change of basis (nothing is shown in the last column in this case) or a change of basis in only one of the subspaces. In few cases, the transformation matrix consists of two matrices different from the identity.
\begin{longtable}{|p{2.0cm}|p{1.7cm}|p{5.0cm}|p{8cm}|}
\caption{\label{tab:equinoncoplanar}List of SpPGs after the addition of the TR operation to the 598 non-coplanar spin point groups. The first column gives the number of the non-coplanar SpPGs with TR. Each block of the second column shows the numbers of the non-coplanar SpPGs that result in the same SpPG after the addition of TR. The symbols of the SpPGs of the second column are included in the third column. The first row in each block of the fourth column gives the notation of the SpPG with TR, which has been constructed from the symbol of the first SpPG (chosen as representative) in the third column. The rest of rows in the block give a transformation matrix ($P^R$,$P^U$), in the ${\bf a},{\bf b},{\bf c}$ and ${\bf a}_s,{\bf b}_s,{\bf c}_s$ notation according to equation (\ref{eq:changebasis}), only if at lest one of the two matrices $P^R$ or $P^U$ is different from the identity. The transformation matrix given transforms the symmetry operations of the corresponding SpPG with TR to the set of operations of the SpPG with TR chosen as representative in the first row. Both matrices are different from the identity only in few cases.}\\
\hline
{\bf N. SpPG}&{\bf N.SpPG}&{\bf Label SpPG}&{\bf Label of SpPG with TR}\\
{\bf  with TR}&&&{\bf and transformation matrix}\\
\hline
1.1.1'&1.1&$^{1}1$&$^{1}1.\textrm{1'}$\\
\hline
2.2.1'&2.2&$^{1}\overline{1}$&$^{1}\overline{1}.\textrm{1'}$\\
&2.5&$^{\overline{1}}\overline{1}$&\\
\hline
2.3.1'&2.3&$^{2}\overline{1}$&$^{2}\overline{1}.\textrm{1'}$\\
&2.4&$^{m}\overline{1}$&\\
\hline
\end{longtable}

The way to access the complete list of symmetry operations of these groups in \emph{SPGENPOS} follows the procedure explained in section \ref{sec:noncoplanar}. The output includes the complete list of symmetry operations: first it gives the operations whose spin component $U$ is a proper operation and then the rest of symmetry operations in red color.

\item {\bf Coplanar SpPGs with TR}

The determination of the coplanar SpPGs with TR has been done following exactly the same procedure. Starting from the 253 coplanar groups, after the addition of TR to all of them, we have checked which sets of symmetry operations are equivalent through the relation given by equation (\ref{eq:equivalence}). Table \ref{tab:equicoplanar} summarizes the results of the determination of the 114 coplanar groups with TR. The information given in the table is exactly the same as the information given in Table \ref{tab:equinoncoplanar} for non-coplanar groups.
\begin{longtable}{|p{2.3cm}|p{2.0cm}|p{5.0cm}|p{6.2cm}|}
\caption{\label{tab:equicoplanar}List of SpPGs after the addition of the TR operation to the 253 coplanar spin point groups. The first column gives the number of the coplanar SpPGs with TR. Each block of the second column shows the numbers of the coplanar SpPGs that result in the same SpPG after the addition of TR. The symbols of the SpPGs of the second column are included in the third column. The first row in each block of the fourth column gives the notation of the SpPG with TR, which has been constructed from the symbol of the first SpPG (chosen as representative) in the third column. The rest of rows in the block give a transformation matrix ($P^R$,$P^U$), in the ${\bf a},{\bf b},{\bf c}$ and ${\bf a}_s,{\bf b}_s,{\bf c}_s$ notation according to equation (\ref{eq:changebasis}), only if at lest one of the two matrices $P^R$ or $P^U$ is different from the identity. The transformation matrix given transforms the symmetry operations of the corresponding SpPG with TR to the set of operations of the SpPG with TR chosen as representative in the first row. Both matrices are different from the identity only in few cases.}\\
\hline
{\bf N.SpPG}&{\bf N.SpPG}&{\bf Label SpPG}&{\bf Label of SpPG with TR}\\
{\bf  with TR}&&&{\bf and transformation matrix}\\
\hline
1.1.1.P1'&1.1.1.P&$^{1}1.\textrm{P}$&$^{1}1.\textrm{P1'}$\\
\hline
2.2.2.P1'&2.2.2.P&$^{1}\overline{1}.\textrm{P}$&$^{1}\overline{1}.\textrm{P1'}$\\
&2.3.3.P&$^{2}\overline{1}.\textrm{P}$&\\
\hline
2.3.4.P1'&2.3.4.P&$^{2_{100}}\overline{1}.\textrm{P}$&$^{2_{100}}\overline{1}.\textrm{P1'}$\\
\hline
3.6.5.P1'&3.6.5.P&$^{1}2.\textrm{P}$&$^{1}2.\textrm{P1'}$\\
\hline
\end{longtable}

The way to access the complete list of symmetry operations of these groups in \emph{SPGENPOS} follows the procedure explained in section \ref{sec:noncoplanar}. The output includes the complete list of symmetry operations: first it gives the operations whose spin component $U$ is a proper operation and then the anti-unitary operations in red color.

\item {\bf Collinear SpPGs with TR}

The determination of the collinear groups with TR is trivial. All the operations of the non-trivial group of any collinear group have the form $\{\pm1||R\}$. If TR is added to this group, each operation $R$ of the orbital space has assigned two operations $\{1||R\}$ and $\{-1||R\}$ and these two operations multiplied by all the operations of the trivial group $^{\infty m}1$. Therefore, the whole group can be obtained as the direct product of a trivial group with all $U$ components $U=1$ and the operations of the spin-only group $^{\infty/m m}1$. There are only 32 different trivial groups with all $U=1$, one for each parent group, equivalent to the 32 gray magnetic point groups.

The way to access the complete list of symmetry operations of these 32 groups in \emph{SPGENPOS} follows the procedure explained in section \ref{sec:noncoplanar}. As the output of collinear groups without TR, the list of operations is displayed in two subsets: first it is included the list of operations of the non-trivial group and, in a second subset, the operations of the spin-only group are displayed in parametric form for the $U$ components. It consists of four matrices that represent: (1) a rotation of angle $\varphi$ around the $w$ axis, (2) a 2-fold axis perpendicular to $w$ and forming an angle $\varphi$ with respect to the $u$ direction, (3) a roto-inversion axis which is the product of the axis (1) and TR, and (4) a mirror plane as product of the 2-fold axis (2) and TR. The last two operations are anti-unitary and are displayed in red color.
\end{enumerate}

 \section{Tabulation of the spin-splitting of energy bands allowed by the SpPGs to lowest order}
\label{sec:spinsplit}In this section we present the tables of allowed effective Zeeman field ${\bf B}({\bf k})$ to lowest order in the expansion series given by equation (\ref{eq:Bfield}) of the main text for all non-coplanar and coplanar SpPGs with and without TR. The setting used for each SpPG has been taken from the tool \emph{SPGENPOS} hosted in the Bilbao Crystallographic Server and whose details have been described in section \ref{sec:spgenpos}. The calculations have been performed using the link to \emph{STENSOR}, also hosted in the Bilbao Crystallographic Server, in the output page of \emph{SPGENPOS} once a specific SpPG has been chosen. In some cases, when the lowest non-zero term allowed by the SpPG is $\ell=7,9$, the long calculation time has required a local version of \emph{STENSOR}.

Tables \ref{spinsplit-non}, \ref{spinsplit-cop}, \ref{spinsplit-nonTR} and \ref{spinsplit-copTR} reproduce the results for SpPGs of type non-coplanar, coplanar, non-coplanar with TR and coplanar with TR, respectively. The format of the tables is exactly the same, except in the last table (coplanar SpPGs with TR). The first and second columns give the number and symbol of the SpPGs, respectively, according to the conventions introduced in \emph{SPGENPOS} and detailed in section \ref{sec:spgenpos}. The third column indicates the wave-type of the allowed lowest order in the expansion of equation (\ref{eq:Bfield}) in the main text. The $x$ symbol refers to the non-spin-split pure antiferromagnets ({\bf B}=0). The rest of symbols correspond to the usual wave-type labels, \mbox{$\ell=0\rightarrow s$}, \mbox{$\ell=1\rightarrow p$}, \mbox{$\ell=2\rightarrow d$}, \mbox{$\ell=3\rightarrow f$}, \mbox{$\ell=4\rightarrow g$}, \mbox{$\ell=5\rightarrow h$}, \mbox{$\ell=6\rightarrow i$}, \mbox{$\ell=7\rightarrow j$}, \mbox{$\ell=8\rightarrow k$} and \mbox{$\ell=9\rightarrow \ell$}. The fourth column gives the allowed independent lowest order terms in the expansion series of equation (\ref{eq:Bfield}) in the main text. The general form of the effective Zeeman field ${\bf B}({\bf k})$ is a linear combination of these independent terms with arbitrary coefficients. The format of the Table \ref{spinsplit-copTR} is slightly different. Being the SpPGs coplanar groups with TR, as discussed in the main text, in these groups the restriction $B_u=B_v=0$ is always fulfilled. Therefore, the fourth column of the Table \ref{spinsplit-copTR} indicates only the $B_w$ component of the field.
\begin{center}
\begin{longtable}{|p{2.3cm}|p{4.0cm}|p{0.7cm}|p{7.2cm}|}
		\caption{\label{spinsplit-non}Allowed lowest order term in the expansion series of the effective Zeeman field (equation (\ref{eq:Bfield}) in the main text) for non-coplanar SpPGs. The first and second columns indicate the number and symbol of the group. The third column gives the wave-type. The fourth column shows the allowed independent terms of lowest order. Each row represents an independent term of the same order in the expansion given by equation \ref{eq:Bfield} in the main text.}\\
		\hline
		\multicolumn{2}{|c|}{SpPG}&type&$(B_u,B_v,B_w)$\\
		\hline
1.1&$^{1}1$&s&$(1,0,0)$\\
&&&$(0,1,0)$\\
&&&$(0,0,1)$\\
\hline
2.2&$^{1}\overline{1}$&s&$(1,0,0)$\\
&&&$(0,1,0)$\\
&&&$(0,0,1)$\\
\hline
 \end{longtable}\end{center}
\begin{center}\begin{longtable}{|p{2.3cm}|p{4.0cm}|p{0.7cm}|p{7.2cm}|}
		\caption{\label{spinsplit-nonTR}Allowed lowest order term in the expansion series of the effective Zeeman field (equation (\ref{eq:Bfield}) in the main text) for non-coplanar SpPGs with TR. The first and second columns indicate the number and symbol of the group. The third column gives the wave-type. The fourth column shows the allowed independent terms of lowest order. Each row represents an independent term of the same order in the expansion given by equation \ref{eq:Bfield} in the main text.}\\
		\hline
		\multicolumn{2}{|c|}{SpPG}&type&$(B_u,B_v,B_w)$\\
		\hline
1.1.1'&$^{1}1.1'$&p&$k_1(1,0,0)$\\
&&&$k_1(0,1,0)$\\
&&&$k_1(0,0,1)$\\
&&&$k_2(1,0,0)$\\
&&&$k_2(0,1,0)$\\
&&&$k_2(0,0,1)$\\
\hline \end{longtable}\end{center}
\begin{center}\begin{longtable}{|p{2.3cm}|p{4.0cm}|p{0.7cm}|p{7.2cm}|}
		\caption{\label{spinsplit-cop}Allowed lowest order term in the expansion series of the effective Zeeman field (equation (\ref{eq:Bfield}) in the main text) for coplanar SpPGs. The first and second columns indicate the number and symbol of the group. The third column gives the wave-type. The fourth column shows the allowed independent terms of lowest order. Each row represents an independent term of the same order in the expansion given by equation \ref{eq:Bfield} in the main text.}\\
		\hline
		\multicolumn{2}{|c|}{SpPG}&type&$(B_u,B_v,B_w)$\\
		\hline
1.1.1.P&$^{1}1.P$&s&$(1,0,0)$\\
&&&$(0,1,0)$\\
\hline
2.2.2.P&$^{1}\overline{1}.P$&s&$(1,0,0)$\\
&&&$(0,1,0)$\\
\hline
2.3.3.P&$^{2}\overline{1}.P$&x&\\
\hline
2.3.4.P&$^{2_{100}}\overline{1}.P$&s&$(1,0,0)$\\
\hline
 \end{longtable}\end{center}
\begin{center}\begin{longtable}{|p{2.3cm}|p{5.0cm}|p{0.7cm}|p{6.2cm}|}
		\caption{\label{spinsplit-copTR}Allowed lowest order term in the expansion series of the effective Zeeman field (equation (\ref{eq:Bfield}) in the main text) for coplanar SpPGs with TR. The first and second columns indicate the number and symbol of the group. The third column gives the wave-type. The fourth column shows the allowed independent terms of lowest order.  Each row represents an independent term of the same order in the expansion given by equation \ref{eq:Bfield} in the main text. As $B_u=B_v=0$ in all the case, only the component $B_w$ is explicitly given.}\\
		\hline
		\multicolumn{2}{|c|}{SpPG}&type&$B_w$\\
		\hline
1.1.1.P1'&$^{1}1.P1'$&p&$k_1$\\
&&&$k_2$\\
&&&$k_3$\\
\hline
2.2.2.P1'&$^{1}\overline{1}.P1'$&x&\\
\hline
2.3.4.P1'&$^{2_{100}}\overline{1}.P1'$&p&$k_1$\\
&&&$k_2$\\
\hline
 \end{longtable}\end{center}

\bibliography{references}

\end{document}